\documentclass[prl,twocolumn,superscriptaddress]{revtex4-1}
\usepackage{amssymb,amsmath,amsthm,amsfonts,graphicx}
\usepackage{latexsym}
\usepackage{bm}
\usepackage[]{hyperref}
\usepackage{tensor}
\usepackage[scr=euler,calscaled=.96]{mathalfa}
\usepackage{dcolumn}
\usepackage[utf8]{inputenc}
\allowdisplaybreaks

\pdfoutput=1

\begin{document}
	
\title{Plausible scenario for a generic violation of the weak cosmic censorship conjecture in asymptotically flat four dimensions}

\author{Felicity~C.~Eperon}
\email{fce21@cam.ac.uk}
\affiliation{Department of Applied Mathematics and Theoretical Physics, University of Cambridge, Wilberforce Road, Cambridge CB3 0WA, UK} 
\author{Bogdan~Ganchev}
\email{bvg25@cam.ac.uk}
\affiliation{Department of Applied Mathematics and Theoretical Physics, University of Cambridge, Wilberforce Road, Cambridge CB3 0WA, UK} 
\author{Jorge~E.~Santos}
\email{jss55@cam.ac.uk}
\affiliation{Department of Applied Mathematics and Theoretical Physics, University of Cambridge, Wilberforce Road, Cambridge CB3 0WA, UK} 

\begin{abstract}
	We present a plausible counterexample to the weak cosmic censorship conjecture in the four-dimensional Einstein-Scalar theory with asymptotically flat boundary conditions. Our setup stems from the analysis of the massive Klein-Gordon equation on a fixed Kerr black hole background. In particular, we construct the quasinormal spectrum numerically, and analytically in the WKB approximation, then go on to compute its backreaction on the Kerr geometry. In the regime of parameters where the analytic and numerical techniques overlap we find perfect agreement. We give strong evidence for the existence of a nonlinear instability at late times.
\end{abstract}

\maketitle
\textbf{~Introduction --} Quantum gravity remains \emph{terra incognita}, largely because it is hard to access experimentally. One might wonder why, since singularities are known to form under a variety of circumstances \cite{Senovilla:2018aav}. However, general relativity stubbornly conceals these regions of high curvature, where quantum gravity is likely to play a leading role, by hiding them behind an absolute event horizon. This phenomenon is observed for such large classes of initial data that it was promoted to a conjecture in \cite{Penrose:1969pc}. This is the weak cosmic censorship conjecture (WCCC), which forbids the formation of naked singularities, \emph{i.e.} singularities in causal contact with future null infinity, starting from generic initial data.

Here we propose a four-dimensional asymptotically flat counterexample to the WCCC motivated by the \emph{superradiant instability} \cite{Bardeen:1972fi,Starobinsky:1973aij,Damour:1976kh,Zouros:1979iw,Detweiler:1980uk,Dolan:2007mj,Yoshino:2013ofa,Brito:2014wla,Brito:2015oca,Shlapentokh-Rothman:2013ysa}, afflicting massive scalar perturbations around Kerr black holes (BHs) \cite{Kerr:1963ud} - rotating, with spherical topology and considered the most general BH solutions of the vacuum Einstein equation \cite{Robinson:2004zz}.

The angular dependence of such perturbations around Kerr is parametrized by two integers $\{m,\ell\}$: $m$ counts the number of nodes in the azimuthal direction and $\ell-|m|$ the number of zeros in the polar direction. We show that for \emph{any} scalar field of mass $\mu$ and \emph{any} nonzero value of the BH spin, for sufficiently large values of $\ell=m$,  these perturbations herald instabilities around the BH, extracting energy and angular momentum. Furthermore, the time scales associated to these instabilities grow parametrically as $e^{4\,\ell\,\log\ell}$, indicating that each of the $\ell$ modes decouples from the rest, evolving independently. 

As time progresses, modes with smaller values of $\ell$ stabilize one by one, forming scalar clouds around the BH, similar to those in \cite{Chodosh:2015oma,Herdeiro:2014goa}. However, these BHs were shown to be unstable to higher $m$-modes \cite{Ganchev:2017uuo}, giving rise to the expectation of a cascade towards larger values of $\ell$. This corresponds to a transfer of energy from lower $\ell$-modes to higher ones, indicating an evolution towards smaller scales - a phenomenon akin to turbulence in nonrelativistic $3+1$ fluids.

A possible stabilizing mechanism is the emission of gravitational waves (GWs) by the scalar clouds \cite{Sasaki:2003xr,Brito:2017zvb}. Were they to dissipate energy faster than superradiance creates them, the above scenario would not be possible. We numerically compute the GW emission for fixed gravitational coupling, $M\,\mu$, and spin parameter, $J/M^2$, as a function of $\ell=m$ and find that it leads to energy and angular momentum dispersion that would not be able to counter the efficiency of superradiance.

Our paper is organized as follows: first we present our setup and provide analytic and numerical data for the instability time scales at large $\ell$. We then compute, numerically, the energy radiated towards future null infinity in this process as well as the backreaction of the scalar on one of the components of the Weyl tensor. We see that modes with higher $\ell$ radiate less, implying that energy is accumulated at small scales more efficiently for larger values of $\ell$. Finally we end with discussion of the results.

\textbf{~Setup --} We work with the Einstein-Hilbert action minimally coupled to a real massive scalar field $\psi$
\begin{equation}
S=\int_{\mathcal{M}} \mathrm{d}^4x\sqrt{-g}\left(\frac{R}{16\,\pi\,G}-\nabla_a\psi\nabla^a\psi-\mu^2\psi^2\right),
\end{equation}
where $\mu$ is the scalar field mass, $g_{ab}$ the spacetime metric and $R$ its Ricci scalar. The equations of motion are
\begin{subequations}
	\begin{equation}
	R_{ab}-\frac{R}{2}g_{ab}= 8\pi\,G\,T_{ab}
	\label{eq:einstein}
	\end{equation}
	where $R_{ab}$ is the Ricci tensor of $g_{ab}$,
	\begin{equation}
	\Box \psi =\mu^2 \psi\,,
	\label{eq:KGeq}
	\end{equation}
	and
	\begin{equation}
	T_{ab}=2\nabla_a \psi\nabla_b \psi-g_{ab}\nabla^c \psi\nabla_c \psi-\mu^2\psi^2 g_{ab}\,.
	\label{eq:stress}
	\end{equation}
	\label{eqs:EOM}
\end{subequations}
An important solution to these equations is the Kerr BH \cite{Kerr:1963ud}, with $\psi=0$ and
\begin{multline}
\mathrm{d}s^2 = -\frac{\Delta}{\Sigma^2}\left(\mathrm{d}t-a \sin^2\theta \mathrm{d}\phi\right)^2+\Sigma^2\left(\mathrm{d}\theta^2+\frac{\mathrm{d}r^2}{\Delta}\right)
\\
+\frac{\sin^2\theta}{\Sigma^2}[a\,\mathrm{d}t-(r^2+a^2)\mathrm{d}\phi]^2\,,
\label{eq:kerr}
\end{multline}
where $\Delta = r^2+a^2-2\,M\,r$, $\Sigma^2=r^2+a^2\cos^2\theta$, $\phi\in(0,2\pi)$ is a periodic coordinate and $\theta\in(0,\pi)$ is a polar coordinate.  The BH event horizon is a null hypersurface with $r=r_+\equiv M+\sqrt{M^2-a^2}$, angular velocity $\Omega_K = a/(a^2+r_+^2)$ and surface gravity $\kappa_K=(r_+^2-a^2)/[2 r_+(r_+^2+a^2)]$. The constant $M$ is the BH mass and $a$ parametrizes its angular momentum via $J = M\,a$. The absence of naked singularities demands $|a|\leq M$ with the inequality being saturated at extremality, when the Kerr BH event horizon becomes degenerate with $\kappa_K=0$.

We study Eq.~(\ref{eq:KGeq}) on a fixed Kerr background (\ref{eq:kerr}), which is stationary and axisymmetric with respect to the Killing fields $\partial /\partial t$ and $\partial/\partial \phi$, respectively. We consider perturbations of the form
\begin{equation}
\psi(t,r,\theta,\phi) = e^{-i\,\omega\,t+i\,m\,\phi}\widehat{\psi}_{\omega m}(r,\theta)
\label{eq:KGAnsatz}
\end{equation}
and assume that $\widehat{\psi}_{\omega m}(r,\theta)$ is separable, $\widehat{\psi}_{\omega m}(r,\theta)=R_{\omega \ell m}(r)S_{\omega \ell m}(\theta)$, with the label $\ell$ anticipating that the separation constant will be parametrized by an integer $\ell$. Not all solutions to Eq.~(\ref{eq:KGeq}) are separable, but we are interested in those composed of the sum (possibly infinite) of such separable solutions. $\omega\in\mathbb{C}$ is a complex frequency, determined by imposing appropriate boundary conditions. We are interested in finding unstable mode solutions for which $\mathrm{Im}(\omega)>0$.

Inserting the \emph{ansatz} (\ref{eq:KGAnsatz}) into Eq.~(\ref{eq:KGeq}) yields a system of two second order ordinary differential equations, coupled via the separation constant $\Lambda$,
\begin{subequations}
	\begin{equation}
	\Delta\,\big[\Delta\,{R_{\omega \ell m}}_{,r}\big]_{,r}+V(r)R_{\omega \ell m}=0,\label{eq:radialr}
	\end{equation}
	\begin{multline}
	\frac{1}{\sin\theta}\big[\sin\theta\,{S_{\omega \ell m}}_{,\theta}\big]_{,\theta}
	\\
	-\left[a^2\,k^2\cos^2\theta+\frac{m^2}{\sin^2\theta}-\Lambda\right]S_{\omega \ell m}=0,
	\label{eq:ang}
	\end{multline}
	where
	\begin{align}
	V(r)&=-k^2r^4+2M \mu ^2 r^3-(\Lambda+a^2 k^2)r^2+\notag\\
	&(2M\Lambda -4a m M \omega+2Ma^2 \omega^2)r-a^2(\Lambda-m^2)\,,
	\end{align}
\end{subequations}
with $k\equiv \sqrt{\mu^2-\omega^2}$. Finding unstable modes amounts to finding the values of $\omega$ for which $\psi$ has ingoing boundary conditions at the event horizon (consistent with the equivalence principle) and finite energy on a partial Cauchy surface $t={\rm const}$.

This problem can be tackled numerically (for any values of the parameters) and analytically (in certain regions of moduli space). We will first compute the modes using a WKB expansion in $m$, which we detail next.

\textbf{~WKB Expansion and Numerical Validation--} Our WKB expansion is valid for any spin parameter $|a|<M$, and only assumes $m$ to be large. For small values of $a$ and $\mu$, it reproduces the results in \cite{Detweiler:1980uk}\footnote{The large $m$ limit does not commute with the extremal limit. It can be shown that at any finite value of $m$, $\mathrm{Im}(\omega)=0$ at extremality. However, the region of the black hole moduli space where this behaviour occurs scales inversely with $m$, and is thus absent in the strict $m\to+\infty$ limit. The behaviour in this region could be investigated by switching the order of limits.} (up to a factor of $2$, see \cite{Furuhashi:2004jk}). For large $m$, Eq.~\eqref{eq:ang} can be approximated by the usual equation for spherical harmonics on a 2-sphere so that $\Lambda=\ell(\ell+1)+O(\ell^{-1})$ and $\ell,\,m$ are integers with $\ell\geq0$, $|m|\leq \ell$. In the following we set $m=\ell$ and take the limit $\ell\gg 1$.

To determine the large $\ell$ limit at fixed $\mu$ and $a$, we used the same method as in \cite{Eperon:2016cdd}, which combines a matched asymptotic and WKB type approach. We quote the final result and leave the details to the Supplemental Material,
\begin{subequations}
	\begin{equation}
	\mathrm{Re}(\omega M)=\hat{\mu}\left(1-\frac{\hat{\mu}^2}{2\ell^2}\right)+\mathcal{O}(\ell^{-3})\,,
	\end{equation}
	\begin{multline}
	\mathrm{Im}(\omega M)=\frac{\ell^{-4 \ell-\frac{9}{2}+p }}{2^{2 \ell+1-p }\sqrt{\pi } p ! } \hat{\mu}^{4 \ell+5} \sinh \left[\frac{\pi  \left(\ell \Omega _K-\mu \right)}{\kappa _K}\right]\times
	\\
	\exp \left[-\frac{2}{\kappa _K}\left(\ell \Omega _K-\frac{\hat{\mu}}{r_+}\right) \arctan \left(\frac{\Omega_K}{\kappa_K}\right)-2 (1-\ell+p )\right]\times
	\\
	\left[1+\mathcal{O}(\ell^{-1})\right]\,,
	\label{eq:freIm}
	\end{multline}
\end{subequations}
where $\hat{\mu}\equiv \mu M$ and $p\in\mathbb{N}_0$ is a radial overtone. First note that if we set $a=0$, the argument of the $\sinh$ becomes negative, and the instability disappears. Furthermore, its onset sits precisely at the onset of superradiance, namely $\ell\Omega_K=\mathrm{Re}(\omega)$. More importantly for our purposes, in the limit $\ell\to+\infty$, the growth rate of the instability scales as $e^{-4 \ell \log \ell}$, and no matter what the value of $\mu$ or $a$, one can always find a value of $\ell=\ell_{\star}\equiv\lceil\mu/\Omega_K\rceil$ above which the instability sets in. This shows that \emph{all} Kerr black holes are unstable to massive scalar field perturbations, irrespective of their initial spin $|J|<M^2$ and of the mass $\mu$ of the scalar perturbation.

One can test the regime of validity of our approximation, by comparing with data obtained by solving numerically, without approximations, the full equations. To this end, we perform a change of variables so that we only solve for smooth functions in their integration domain. This necessarily involves a choice of normalization, which we describe next. For numerical aid we define
\begin{subequations}
	\begin{equation}
	R_{\omega \ell m}(r) = \left(1-\frac{r_+}{r}\right)^{-i \frac{\omega-m \Omega_K}{2 \kappa_K}}e^{-k\,r}\left(\frac{r_+}{r}\right)^\gamma q_r(r)\,,
	\end{equation}
	\begin{equation}
	S_{\omega \ell m}(\theta) = \sin^{m}\theta\;q_\theta(\theta)\,,
	\end{equation}
	\label{eqs:num}%
\end{subequations}
with $q_r(r_+)=q_\theta(0)=1$ and $\gamma \equiv 1+2 M k-M \mu ^2/k$. We now use the methods of \cite{Dias:2015nua} to solve for the eigenpair $(\omega,\Lambda)$ using a Newton-Raphson routine \footnote{The only additional complication is that we are searching for extremely small growth rates when $m$ increases, so using extended precision is mandatory.}.

\begin{figure}
	\centering
	\includegraphics[width=.47\textwidth]{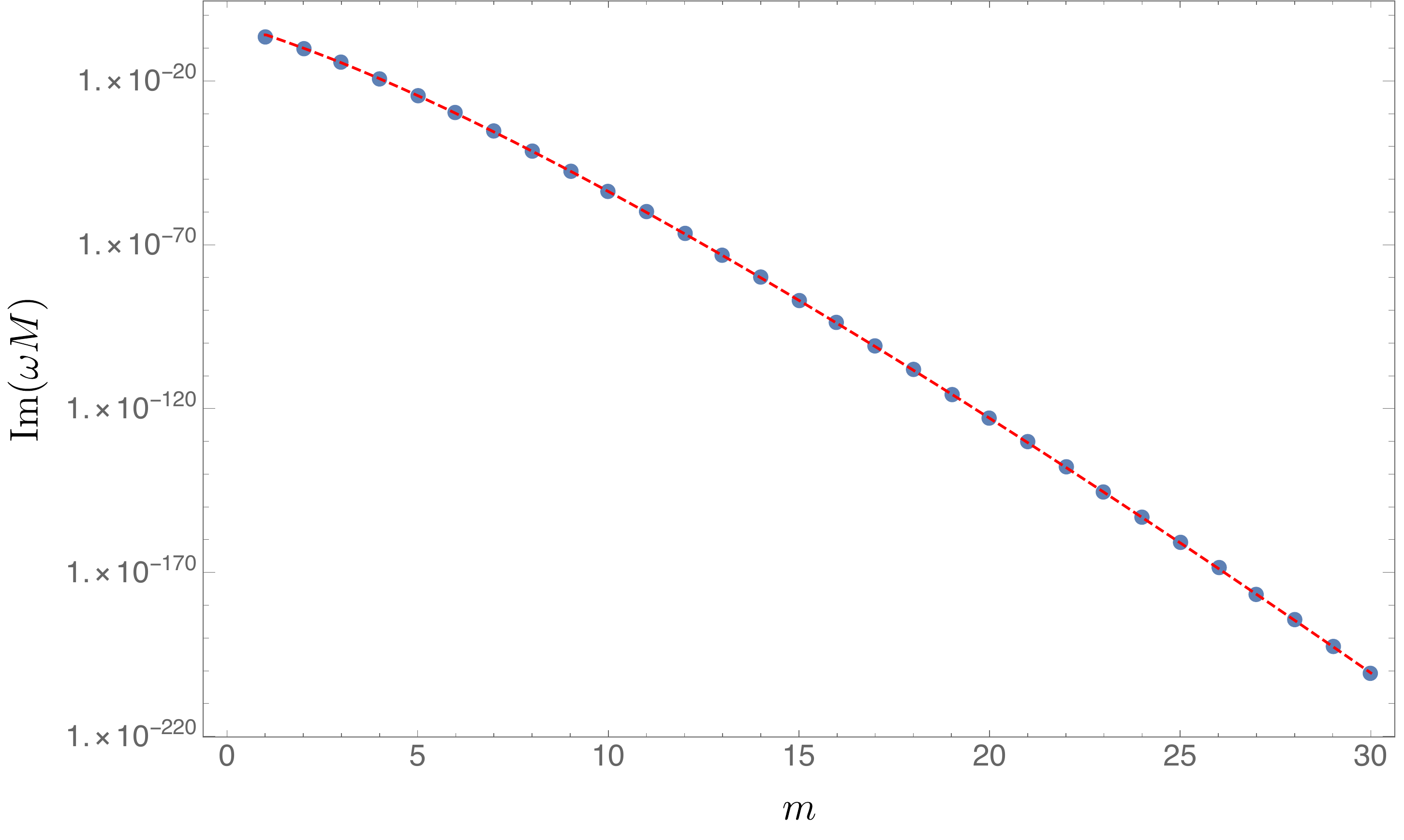}
	\caption{The superradiant modes of a massive scalar around Kerr with $M\,\mu=0.42$ and $J/M^2=0.99$, as a function of $m$. The dashed red curve shows the analytic expression (\ref{eq:freIm}) and the blue disks our exact numerical data.}
	\label{fig:freqs}
\end{figure}

As seen in Fig.~\ref{fig:freqs}, our numerical data agrees excellently with \eqref{eq:freIm}. Furthermore, one can measure deviations of our WKB expression to the exact numerical result, and it agrees with the error given in Eq.~(\ref{eq:freIm}).

\textbf{~Backreaction --} We want the GWs emitted by a scalar cloud around a Kerr BH and its leading order backreaction on the geometry. In the vector field case \cite{East:2017ovw,East:2017mrj} it has been shown that the system evolves adiabatically; the emergence of the cloud due to superradiance, and the consecutive saturation of the vector mode responsible, due to the spinning down of the BH, proceed on a much faster time scale than the dispersion of energy and angular momentum due to GW emission from the cloud.

We proceed using nonlinear perturbation theory and declare
\begin{equation}
\psi=\sum_{i=0}^{+\infty} \psi^{(2i+1)}\varepsilon^{2i+1}\,,\quad\text{and}\quad g=g_{K}+\sum_{i=1}^{+\infty} g^{(2i)}\varepsilon^{2i}\,,
\end{equation}
where $g_{K}$ is given by the Kerr metric (\ref{eq:kerr}). We expand the equations of motion (\ref{eqs:EOM}) in a power series in $\varepsilon$. To first order in $\varepsilon$ we solve eq.~(\ref{eq:KGeq}) subject to a choice of initial data. For the case at hand, we choose $\psi$ to be given by the real part of one of the unstable modes we have determined above. These are labelled by a given value of $m$. Furthermore, since  $\mathrm{Im}(\omega M)\ll \mathrm{Re}(\omega M)$, we take $\omega$ to be purely real. We then proceed to second order and attempt to compute the leading order backreaction on the metric, $g^{(2)}$, and its associated curvature. The standard approach to the linearized Einstein equation presents us with a daunting task; however, Kerr BHs are algebraically special, allowing us to bypass computing $g^{(2)}$, and directly calculate certain gauge invariant scalars built out of the Weyl tensor. These do not couple amongst themselves and we focus on the Newman-Penrose scalar $\psi_4$ since it also allows us to efficiently compute the GWs emitted by the scalar cloud. $\psi_4$ obeys the Teukolsky equation \cite{Teukolsky:1973ha,Press:1973zz,Teukolsky:1974yv}
\begin{align}
\Big[\big(\Delta+3\gamma-\bar{\gamma}+4\mu+\bar{\mu}\big)\big(D+&4\epsilon-\rho\big)-3\psi_2-\notag\\
-\big(\bar{\delta}+3\alpha+\bar{\beta}+4\pi-\bar{\tau}\big)\big(\delta+&4\beta-\tau\big)\Big]\psi_4=4\pi\,T_4\,,\label{eq:teukPsi4}
\end{align}
whereby all quantities appearing in Eq.~(\ref{eq:teukPsi4}) are given in the Supplemental Material.  The source term $T_4$ is reconstructed directly from Eq.~(\ref{eq:stress}) and thus depends on $\psi^{(1)}$ and its gradient only.

The lhs of \eqref{eq:teukPsi4} can be separated into angular and radial parts as in the vacuum case \cite{Teukolsky:1973ha}, allowing us to solve for $\psi_4$ as an infinite sum of separable solutions using Green's method.

From $\psi_4$, Teukolsky \cite{Teukolsky:1974yv} showed us how to compute the rate of gravitational radiation at future null infinity
\begin{align}
\frac{\mathrm{d}^2E_{s}}{\mathrm{d}t\mathrm{d}\Omega}=\lim\limits_{r\rightarrow\infty}\frac{r^2}{4\,\pi\,\hat{\omega}^2}\left\lvert\psi_4\right\rvert^2,
\label{eq:energyRate}
\end{align}
where $\hat{\omega}=2\,\omega$ and $\mathrm{d}\Omega$ is the induced volume on a unit 2-sphere.

We work with the scaled expression
\begin{align}
P_{E}=\frac{\mathrm{d}E_{s}}{\mathrm{d}t}\left(\frac{M}{\mathcal{M}_s}\right)^2\,,
\label{eq:energyRateNum}
\end{align}
where
\begin{equation}
\mathcal{M}_s=\int_{r_+}^{+\infty}\int_{0}^{\pi}\int_{0}^{2\pi}\sqrt{-g}\,\tensor{T}{^t_t}\,\mathrm{d}\phi\,\mathrm{d}\theta\,\mathrm{d}r
\end{equation}
is the total scalar field energy, - the energy of the perturbed initial data. $P_{E}$ is independent of the scalar field amplitude and measures the energy radiated per $m$ mode in the initial data. See the Supplemental Material on how to compute $P_E$.

As a measure of curvature, we monitored the following time independent quantity as a function of $m$:
\begin{equation}
\chi \equiv \max_{r,\theta}\;\left(|\psi_4|^2/\mathcal{M}_{s}^2\right).
\label{eq:max}
\end{equation}
The radial and azimuthal location of the maximum of (\ref{eq:max}), $(r,\theta)=(r_{\star},\theta_\star)$, track the maximum of $[\psi^{(1)}]^2$.

\begin{figure}
	\centering
	\includegraphics[width=.47\textwidth]{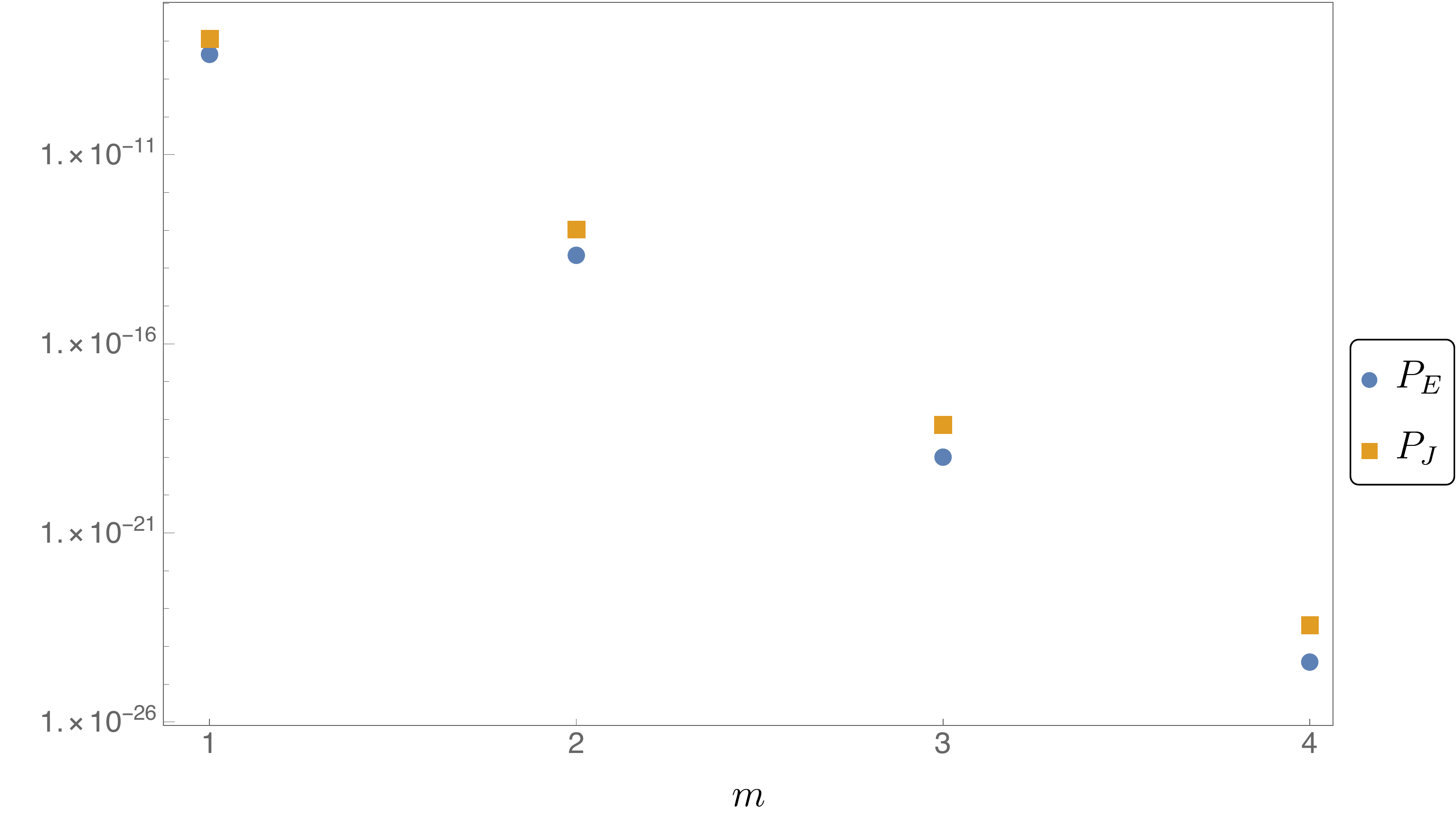}
	\caption{GW emission of energy and angular momentum, $P_E$ and $P_J$ respectively, for a single $\ell=m$ scalar cloud around Kerr as a function of $m$. Same parameters as in Fig.~\ref{fig:freqs}.}
	\label{fig:GWemission}
\end{figure}

Our results for the GW emission are shown in Fig.~\ref{fig:GWemission}. The radiated angular momenta in this process is $P_{J}=\frac{m}{\mathrm{Re}(\omega M)}P_{E}$, in accordance with \cite{Teukolsky:1974yv}. The fact that both $P_{E}$ and $P_{J}$ appear to decrease rapidly with increasing $m$ shows that the evolution occurs, to very good approximation, at fixed energy and angular momentum. This is akin to the time evolution of the superradiance instability with anti-de Sitter asymptotics \cite{Dias:2011at,Niehoff:2015oga}, simulated recently in \cite{Chesler:2018txn}, and shows hints of turbulent behaviour.

The data in Fig.~(\ref{fig:GWemission}) is for a fixed value of the dimensionless spin parameter $a/M$. However, during the aforementioned cascade the BH will be gradually spinning down, hence the gravitational radiation for each value of $m$ should ideally be computed by accounting for BH's loss of energy due to the superradiant modes active prior to the one under consideration. Nevertheless, using the superradiant condition $\mathrm{Re}(\omega)>m \Omega_K$, one sees that $\Delta(a/M)$ for successive superradiant modes $\sim\ell^{-1}$ as $\ell\rightarrow\infty$, implying that in the regime of interest the dimensionless spin will be approximately constant and Fig.~(\ref{fig:GWemission}) represents accurately the qualitative behaviour.

\begin{figure}
	\centering
	\includegraphics[width=.47\textwidth]{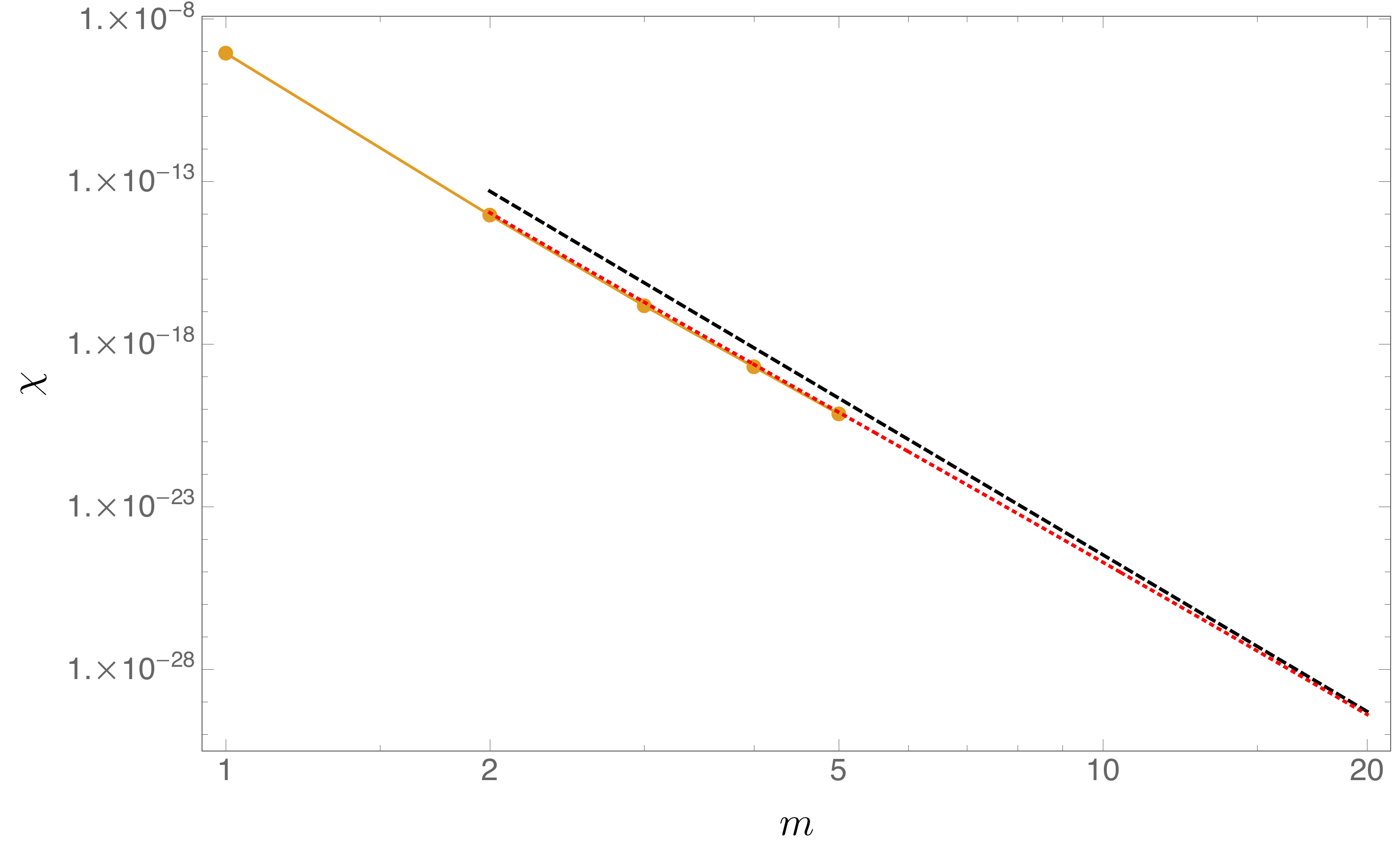}
	\caption{$\chi$, a measure of the spacetime curvature, as a function of $\ell=m$. Same parameters as in Fig.~\ref{fig:freqs}. The black dashed curve shows our leading order approximation, whereas the dotted red line includes the next to leading order correction and the orange circles are numerical data.}
	\label{fig:back}
\end{figure}

In Fig.~\ref{fig:back} we show the dependence of $\chi$ on the initial data, here labelled by $m$. A WKB-type analysis reveals that the leading behaviour for $\chi$ is power law in $1/m$
\begin{equation}
\chi_{\mathrm{WKB}} = \frac{\hat{\mu}^{10}}{m^{16}\pi^2 2^6}\left(\frac{\mu}{r_+}\right)^6\,. 
\end{equation}
One can consistently correct this approximation to next-to-leading order (see the Supplemental Material).

Note that $P_E$ is harder to compute numerically than $\chi$, hence why we have extended results for $\chi$ up to $m=5$.

\textbf{\textit{~Gedanken} experiment --} We are now ready to present our possible counterexample to WCCC with asymptotically flat boundary conditions. Consider generic initial data for the Einstein-Scalar system. These data are controlled by a large functional freedom coming from the fact that we can choose the initial metric on a constant time slice, as well as the extrinsic curvature (so long as the Hamiltonian and momentum constraints are satisfied). In addition, we can also control the initial profile for the scalar field and its first time derivative on a constant time slice. We are going to choose our initial data to be close to that of the Kerr BH, so that deviations from the Kerr metric only occur at order $\mathcal{O}(\psi^2)$. This condition can be relaxed by considering initial data for the purely gravitational sector that is small in some norm. This essentially means that all the dynamics are being generated by the scalar field.

For generic scalar field initial data, we expect the scalar field profile to have some support on the unstable modes of the preceding sections, \emph{i.e.} to excite unstable modes. Since all other modes decay with time \footnote{This decay can be very complicated to determine and is not exponential with time.}, we expect the late time evolution to be dominated by the leading unstable modes and their backreaction. For each value of $m$ there is an infinite number of such modes labelled by $\ell\geq|m|$. However, all of these modes stop being unstable as soon as the condition $\mathrm{Re}(\omega)>m \Omega_K$ is no longer satisfied. The dynamics of this change in angular momentum and energy is entirely controlled by the $\ell=m$ modes, implying that, after some suitably long time, the dynamics of the Einstein-Scalar system can be well approximated by restricting our attention to scalar profiles of the form
\begin{equation}
\psi(t,r,\theta,\phi)= \mathrm{Re}\left[\sum_{\ell=0}^{+\infty} a_{\ell}e^{-i\omega_{\ell}+i \ell \phi}\,R_{\omega_{\ell} \ell \ell}(r)\,S_{\omega_{\ell} \ell \ell}(\theta)\right]\,,
\label{eq:appro}
\end{equation}
and determining its leading order backreaction on the spacetime curvature. The coefficients $a_{\ell}$ are determined by our choice of initial data: for finite Sobolev norm initial data we expect $a_{\ell}$ to exhibit polynomial behaviour in $1/\ell$, whereas for $C^{\infty}$ initial data we expect the coefficients $a_{\ell}$ to decay faster than any polynomial in $1/\ell$. Note that for real analytic initial data one can show that $a_{\ell}\approx e^{-\alpha \ell}$, for $\alpha>0$.

Given that each $\ell$ mode evolves on an exponentially different timescale, as shown by Eq.~(\ref{eq:freIm}), they effectively decouple from each other, allowing us to study each term in Eq.~(\ref{eq:appro}) and its backreaction on the metric separately. Eventually, a given $\ell=\ell_{\star}$ mode becomes stable, but the system remains unstable to higher values of $\ell>\ell_{\star}$. This cascading happens slowly, since the time scales for this effect are exponentially large. One might worry that the energy contained in these high $\ell$ modes is radiated away as time passes by, but we have seen in Fig.~(\ref{fig:GWemission}) that this is not the case. In fact, the larger the value of $\ell$, the smaller its radiative power is.

Finally, we have seen in Fig.~(\ref{fig:back}) that $\chi$ decays as $\ell^{-16}$. This in turn implies that a mode with weigh $a_{\ell}$ will depend on $\ell$ as  $a_{\ell}^4\,\ell^{-16}$. The reason for this is simple: the Teukolsky scalar $\psi_4$ is sourced by $[\psi^{(1)}]^2$, and $\chi$ is related to $|\psi_4|^2$, which translates into the overall scaling mentioned above. The curvatures are thus suppressed with time, and the evolution continues until all of the angular momentum is deposited into the scalar cloud, and the central black hole becomes Schwarzschild. However, one can show \cite{futureWork} that massive scalar field perturbations decay extremely slow around Schwarschild black holes, with a dependence as weak as $1/\log(\log t)$, for large $t$. This suggests that the hypothetical end point is itself \emph{nonlinearly unstable} through a mechanism similar to the one reported in \cite{Eperon:2016cdd,Keir:2014oka,Keir:2016azt}. 
On the other hand, this slow decay warrants the question whether the gravitational radiation might not just disperse the clouds, before any nonlinearities become problematic. We cannot answer this in our analysis, as our approximation for radiation emission breaks down at late times. Therefore the outcome of our thought experiment depends on which of the competing processes - the GW emission or the nonlinear issues resulting from slower than logarithmic decay - wins over. If it is the latter, then the lack of a possible stationary end point leads us to conjecture that the spirit, if not the letter, of weak cosmic censorship is violated. Whether the curvature will be infinite in finite time is a question that we cannot settle with our current methods.

Note also that $r_{\star}(\ell)$ increases with $\ell$, posing a problem from a numerical perspective: 1) the time scales involved are enormous; 2) the cascading towards high $\ell$ values makes this problem dependent on high frequency modes (as the simulation of turbulence in $3+1$ nonrelativitic fluids); and 3) the integration domain must extend to spatial infinity to observe this effect.

\textbf{~Conclusions --} We have seen (Fig.~\ref{fig:GWemission}) that the efficiency of superradiance cannot be counteracted by GW emission, implying that the system will continue advancing to higher values of $m$ with curvatures decreasing appropriately (as shown by Fig.~\ref{fig:back}) until a configuration with a central Schwarzschild black hole is reached. However, Schwarzschild is likely to be nonlinearly unstable due to the very slow decay of perturbations induced by massive scalar fields. Reaching this troublesome regime, given that effects we cannot account for in our analysis do not prevent this, will involve time scales much longer than the age of our Universe, of course, as one will have to go to large values of $\ell=m$ . Nevertheless our scenario provides the first plausible example of a system with asymptotically flat boundary conditions, where WCCC is violated.

\begin{acknowledgments}
	\indent{\bf~Acknowledgments --} We would like to thank O.~J.~C.~Dias and H.~S.~Reall for reading an earlier version of this manuscript. We also want to thank W.~E.~East for many useful discussions that led to this paper. F.C.E. and B.G. are supported by STFC. J.E.S is supported in part by STFC Grants PHY-1504541 and ST/P000681/1. This work used the DIRAC Shared Memory Processing system at the University of Cambridge, operated by the COSMOS Project at the Department of Applied Mathematics and Theoretical Physics on behalf of the STFC DiRAC HPC Facility (\href{www.dirac.ac.uk}{www.dirac.ac.uk}). This equipment was funded by BIS National E-infrastructure Capital Grant ST/J005673/1, STFC Capital Grant ST/H008586/1, and STFC DiRAC Operations Grant ST/K00333X/1. DiRAC is part of the National e-Infrastructure.
\end{acknowledgments}

\newpage
\section{Supplemental Material}

\indent{\bf~WKB expansion --} We present the details of the expansion in large $\ell$ that we use to determine the scalar instability growth rate.  We work with $l-m=n$, where $n\sim\mathcal{O}(1)$, $n\in\mathbb{N}_0$. To this end we rewrite the Klein-Gordon equation from the main text in the following way
\begin{subequations}
	\begin{align}
	\frac{r-r_+}{r_+}\big[\Delta\,R_{\omega\ell m}'&(r)\big]_{,r}+V(r)\,R_{\omega\ell m}(r)=0,\label{eq:KGrad}\\
	\frac{\big[\sin\theta\,S_{\omega\ell m}'(\theta)\big]_{,\theta}}{\sin\theta}-&\left[\tilde{a}^2\,\tilde{k}\,\cos^2\theta+\frac{m^2}{\sin^2\theta}-\Lambda\right]S_{\omega\ell m}(\theta)=0,\label{eq:KGang}\\
	V(r)=a_1+&a_2\,\frac{r}{r_+}+a_3\,\frac{r^2}{r_+^2}-\tilde{k}\,\frac{r^3}{r_+^3}+\frac{a_4}{r/r_+-\tilde{a}^2},\notag\\
	a_1=\Lambda+&\tilde{a}\,\tilde{\omega}\left(1+\tilde{a}^2\right)\left[\tilde{a}\,\tilde{\omega}\left(1+\tilde{a}^2\right)-2\,m\right],\notag\\
	a_2=\tilde{a}^2\,\tilde{\omega}^2&\left(1+\tilde{a}^2\right)-\Lambda,\quad a_3=\tilde{\mu}^2+\tilde{a}^2\,\tilde{\omega }^2,\notag\\
	a_4=&\tilde{a}^2\left[m-\tilde{a}\,\tilde{\omega}\left(1+\tilde{a}^2\right)\right]^2,
	\end{align}
\end{subequations}
where $\tilde{k}=\tilde{\mu}^2-\tilde{\omega}^2$, $\Lambda$ is the angular separation constant and we have introduced dimensionsless variables
\begin{align}
\tilde{a}=\frac{a}{r_+},\;\tilde{M}=\frac{M}{r_+},\;\tilde{\omega}=\omega\,r_+,\;\tilde{\mu}=\mu\,r_+,\label{eq:dimlessVars}
\end{align}
As explained in the main text, finding unstable modes amounts to determining the values of $\tilde{\omega}$ for which $\psi$ has ingoing boundary conditions at the event horizon (consistent with the equivalence principle), and finite energy on a partial Cauchy surface $t={\rm const}$ \footnote{For unstable modes, this condition is \emph{equivalent} to requiring the modes to be outgoing at spatial infinity.}.

In the eikonal limit $\ell\gg1$ the spheroidal eigenvalue is $\Lambda=\ell(\ell+1)+O(\ell^{-1})$; this can easily be verified by a series expansion of the angular equation after prefactoring $\sin^m\theta$ from $S(\theta)$. In order to determine $\tilde{\omega}$ we divide the domain in two intersecting regions, solve \eqref{eq:KGrad} in each of them and then match the solutions in the overlap. To this end we introduce the new variable $x=r/r_+-(1+\tilde{a}^2)$
\begin{align}
(x+\tilde{a}^2)\big[\tilde{\Delta}\,R_{\omega\ell m}'(x)\big]_{,x}&+\tilde{V}(x)R_{\omega\ell m}(x)=0,\label{eq:KGradDimless}\\
V(x)=b_1+b_2\,x+b_3\,&x^2-\tilde{k}\,x^3+\frac{a_4}{1+x},\notag\\
b_1=\tilde{\omega}^2\left(1+\tilde{a}^2\right)^2\left(1+4\,\tilde{a}^2\right)&-2\,m\,\tilde{a}\,\tilde{\omega}\left(1+\tilde{a}^2\right)\notag\\
-\tilde{a}^2\big[\ell(&\ell+1)+\left(1+\tilde{a}^2\right)^2\tilde{\mu}^2\big],\notag\\
b_2=-\left(1+\tilde{a}^2\right)\big[\left(1+3\,\tilde{a}^2\right)&\tilde{\mu}^2-3 \left(1+2\,\tilde{a}^2\right)\tilde{\omega}^2\big]\notag\\
-\ell(\ell+&1),\notag\\
b_3=\left(3+4\,\tilde{a}^2\right)\tilde{\omega}^2-&\left(2+3\,\tilde{a}^2\right)\tilde{\mu}^2,\notag
\end{align}
with $\tilde{\Delta}=(1+x)(x+\tilde{a}^2)$

\indent{\it~Near-horizon region --} Using quasimodes \cite{Holzegel:2013kna,Keir:2014oka}, one can show that $\tilde{k}=\mathcal{O}(\ell^{-2})$. We thus take the near-horizon region to be defined by $x\ll \ell^2$ and see that
\begin{align}
\tilde{k}\,x^3\ll x\,m^2,\quad b_3\,x^2\ll x\,m^2,
\end{align}
implying that we can drop the cubic and quadratic terms inside $V(x)$, leaving us with
\begin{equation}
(x+\tilde{a}^2)\big[\tilde{\Delta}\,R_H'(x)\big]_{,x}+\Big[b_1+b_2\,x+\frac{a_4}{1+x}\Big]R_{H}=0.
\end{equation}
The solution \footnote{Multiply the equation by $(1+x)$, then change variables to $u=(x+\tilde{a}^2)/(1-\tilde{a}^2)$.} is a linear combination of Gauss hypergeoemtric functions
\begin{align}
&R_{H}(x)=A_{in}(-u)^{\delta}(1+u)^{\phi}\,\tensor[_2]{F}{_{1}}\Big[c_-,c_+;c;-u\Big]\\
&+A_{out}(-1)^{-2\,\delta}(-u)^{-\delta}(1+u)^{\phi}\,\tensor[_2]{F}{_{1}}\Big[\lambda_-,\lambda_+;2-c;-u\Big],\notag
\end{align}
where $u=(x+\tilde{a}^2)/(1-\tilde{a}^2)$, $A_{in/out}$ are constants and
\begin{align}
\delta=&\frac{i\,\Omega_0}{1-\tilde{a}^2}\bigg[1+\frac{\tilde{a}^4(1-\tilde{a}^4)(2\,\tilde{\mu}^2-3\,\tilde{\omega}^2)}{\Omega_0^2}\bigg]^{\frac{1}{2}},\notag\\
\phi=&-i\,\frac{m-\tilde{a}\,\tilde{\omega}\,(1+\tilde{a}^2)}{1-\tilde{a}^2},\quad\Omega_0=m\,\tilde{a}-\tilde{\omega}\,(1+\tilde{a}^2)\notag\\
c_{\pm}=&\frac{1}{2}\pm\frac{\sqrt{1-4\,b_2}}{2}+\phi+\delta,\quad c=1+2\,\delta,\notag\\
\lambda_{\pm}=&c_{\pm}-c+1.
\end{align}
Demanding ingoing waves only at the horizon ($x=0$) requires setting $A_{out}=0$. To see this, note that in the limit $x\rightarrow0$ the hypergeometric functions take on a constant value to leading order, hence one just needs to know which of $(-u)^{\pm\delta}$ gives the correct behaviour there. This can most easily be deduced by performing a transformation from BL to Kerr coordinates
\begin{align}
\mathrm{d}v=\mathrm{d}t+\frac{r^2+a^2}{\Delta}\mathrm{d}r,\quad \mathrm{d}\chi=\mathrm{d}\phi+\frac{a}{\Delta}\mathrm{d}r.
\end{align}
Looking at the near-horizon limit allows us to obtain
\begin{align}
e^{-i\,\omega\,t}e^{i\,m\,\phi}=e^{-i\,\omega\,v}e^{i\,m\,\chi}(r-r_+)^{i\big(\tilde{\omega}\frac{1+\tilde{a}^2}{1-\tilde{a}^2}-\frac{m\,\tilde{a}}{1-\tilde{a}^2}\big)},
\end{align}
which has opposite sign to what we expect at the horizon, as the transformation eliminates any singular behaviour.

\indent{\it~Far away region --} Before zooming into spatial infinity we make the following transformation to \eqref{eq:KGradDimless}, to ensure that we will get the correct asymptotic behaviour,
\begin{align}
\mathcal{R}_{\omega\ell m}(x)=&h(x)\,Q_{\infty}(x),\,h(x)=\frac{\sqrt{4\,\tilde{k}^2\,x^2+4\,\tilde{k}\,x\,d_2+d_1^2}}{2\,\tilde{k}\,x+d_1},\notag\\
d_1&=\left(1+2\,\tilde{a}^2\right)\left(2\,\tilde{\mu}^2-3\,\tilde{\omega}^2\right),\notag\\
d_2&=\left(1+3\,\tilde{a}^2\right)\tilde{\mu }^2-\left(2+5\,\tilde{a}^2\right)\tilde{\omega}^2.\label{eq:farAwayTransform}
\end{align}
We work with $x\gg1$, which leaves us\footnote{First multiply (\ref{eq:KGradDimless}) by $1+x$, then perform transformation (\ref{eq:farAwayTransform}), followed by dividing the resulting equation by $(1+x)(x+\tilde{a}^2)(t_0+t_1\,x+t_2\,x^2+t_3\,x^3+2\,x^4)/\left[2\,x\,(x+i_0)^2\sqrt{x(x+i_1)+i_0^2}\right]$, with $t_0=i_0^3(1+\tilde{a}^2)-2\,\tilde{a}^2\,i_0^2+\tilde{a}^2\,i_0\,i_1,\quad t_1=i_0(2\,i_1(1+\tilde{a}^2)+2\,\tilde{a}^2)-i_0^2(1+\tilde{a}^2)+2\,i_0^3-\tilde{a}^2\,i_1,\quad t_2=3\,i_0(1+\tilde{a}^2+i_1),\quad t_3=1+\tilde{a}^2+4\,i_0+i_1,\quad i_0=d_1/(2\,\tilde{k}),\quad i_1=d_2/\tilde{k}$, and expand near spatial infinity ($x=1$)} with
\begin{align}
x^2\,Q''_{\infty}(x)+2\,x\,Q'_{\infty}(x)&-(\tilde{k}\,x^2+e_1\,x-b_2)Q_{\infty}(x)=0,\notag\\
e_1=(1+\tilde{a}^2)&(\tilde{\mu}^2-2\,\tilde{\omega}^2),
\end{align}
to solve. The above can be transformed into Whittaker's differential equation with the help of
\begin{align}
x=\frac{z}{2\,\sqrt{\tilde{k}}},\;j(z)=\frac{Q_{\infty}(z)}{z},
\end{align}
resulting in
\begin{align}
j''(z)+\bigg(-\frac{1}{4}-\frac{e_1}{2\,\sqrt{\tilde{k}}\,z}+\frac{b_2}{z^2}\bigg)j(z)=0.\label{eq:farAwayRad}
\end{align}
The general solution to \eqref{eq:farAwayRad} is then a linear combination of the Whittaker functions $W_{k,\nu}(z)$ and $W_{-k,\nu}(-z)$, thus
\begin{align}
R_{\infty}=B_1\,v(z)\,W_{-\kappa,\nu}(-z)+B_2\,v(z)\,W_{\kappa,\nu}(z),\notag\\
\kappa=-\frac{e_1}{2\sqrt{\tilde{k}}},\;\nu=\frac{1}{2}\sqrt{1-4\,b_2},\;v(z)=h(z)/z,
\end{align}
with $B_{1/2}$ constants. Expanding for large $z$, and using
\begin{align}
z=2\,\sqrt{\tilde{k}}\Big[\frac{r}{r_+}-(1+\tilde{a}^2)\Big]\xrightarrow[r\rightarrow\infty]{}2\,\sqrt{\tilde{k}}\frac{r}{r_+},
\end{align}
we see that 
\begin{align}
\lim\limits_{r\rightarrow \infty}\big[B_1\,v(z)\,W_{-\kappa,\nu}(-z)&+B_2\,v(z)\,W_{\kappa,\nu}(z)\big]=\notag\\
B_1\,e^{\sqrt{\tilde{k}}\,r/r_+}r^{-1-\kappa}&+B_2\,e^{-\sqrt{\tilde{k}}\,r/r_+}r^{-1+\kappa}.
\end{align}
Imposing that the solution decays at infinity sets $B_1=0$.

{\it~Matching --} We now perform the matching procedure in the overlapping region $1\ll x\ll \ell^2$. To this end, we take $x\sim\ell^{\frac{3}{2}}$ and expand the near-horizon and far-away solutions for large and small variables respectively.

Looking at the near-horizon region first, we can straightforwardly use the standard asymptotic series of the Hypergeometric function, as all the parameters grow slower than the variable of the function, leaving us with
\begin{align}
\lim\limits_{x\rightarrow\infty}&R_H=A_{in}\frac{(-1)^\delta\,\Gamma[c_+-c_-]\,\Gamma[c]}{\Gamma[c_+]\,\Gamma[c-c_-]}\bigg[\frac{x}{(1-\tilde{a}^2)}\bigg]^{\nu-\frac{1}{2}}\notag\\
&+A_{in}\frac{(-1)^\delta\,\Gamma[c_--c_+]\,\Gamma[c]}{\Gamma[c_-]\,\Gamma[c-c_+]}\bigg[\frac{x}{(1-\tilde{a}^2)}\bigg]^{-\nu-\frac{1}{2}}.\label{eq:NearFar}
\end{align}

Next we consider the expansion of the far-away solution for small variable and again we can use the standard series in the literature to do so, as the parameters outgrow the argument of the Whittaker function, thus
\begin{align}
\lim\limits_{x\rightarrow 0}R_\infty=-B_2&\frac{(2\,\sqrt{\tilde{k}})^{\nu-\frac{1}{2}}\Gamma[-2\,\nu]}{\Gamma[\frac{1}{2}-\nu-\kappa]}x^{\nu-\frac{1}{2}}\notag\\
&-B_2\frac{(2\,\sqrt{\tilde{k}})^{-\nu-\frac{1}{2}}\Gamma[2\,\nu]}{\Gamma[\frac{1}{2}+\nu-\kappa]}x^{-\nu-\frac{1}{2}}.\label{eq:FarNear}
\end{align}

Afterwards we equate the coefficients in front of the equivalent terms in \eqref{eq:NearFar} and \eqref{eq:FarNear} in order to derive
\begin{align}
&\frac{\Gamma[\frac{1}{2}-\nu-\kappa]}{\Gamma[\frac{1}{2}+\nu-\kappa]}=\Big(2\,(1-\tilde{a}^2)\sqrt{\tilde{k}}\Big)^{2\,\nu}G(\ell),\notag\\
&G(\ell)=\frac{\Gamma[c_--c_+]\,\Gamma[-2\,\nu]\,\Gamma[c_+]\,\Gamma[c-c_-]}{\Gamma[2\,\nu]\,\Gamma[c_+-c_-]\,\Gamma[c_-]\,\Gamma[c-c_+]}.\label{eq:matching1}
\end{align}
Furthermore, in the limit $l\rightarrow\infty$ the RHS above has very small real and imaginary parts, implying that $\Gamma[\frac{1}{2}+\nu-\kappa]$ in the denominator on the LHS must have a pole, thus
\begin{align}
\frac{1}{2}+\nu-\kappa=-N,\label{eq:discreteCond}
\end{align}
with $N\in\mathbb{N}_0$, corresponding to the radial node of the scalar field. This allows us to deduce the real part of $\sqrt{\tilde{k}}$
\begin{align}
\operatorname{Re}\big(\sqrt{\tilde{k}}\big)=-\frac{e_1}{2N+2\,\nu+1}.\label{eq:realK}
\end{align}
Next, with $\lvert\tilde{\omega}_R\rvert\gg\lvert\tilde{\omega}_I\rvert$ \cite{Yang:2012he,Dolan:2010wr}, implying $\lvert \tilde{k}_R\rvert\gg\lvert \tilde{k}_I\rvert$, we can expand \eqref{eq:realK} for large $\ell$ to obtain the real part of $\tilde{\omega}$
\begin{align}
\tilde{\omega}_R=\tilde{\mu}-\frac{(1+\tilde{a}^2)^2\,\tilde{\mu}^3}{8\,\ell^2}+\mathcal{O}(\ell^{-3}),\label{eq:realFreq}
\end{align}
confirming our expectations. Moreover, our calculation is accurate to $\mathcal{O}(1)$ only, hence we can replace $\tilde{\omega}$ with $\tilde{\mu}$ everywhere except inside $\tilde{k}$, where the leading order term cancels. This reduces \eqref{eq:matching1} to
\begin{align}
&\frac{\Gamma[\frac{1}{2}-\nu-\kappa]}{\Gamma[\frac{1}{2}+\nu-\kappa]}=\Big[\frac{\tilde{\mu}^2(1-\tilde{a}^4)}{N+\frac{1+\hat{\nu}}{2}}\Big]^{\hat{\nu}}\hat{G}(\ell),\notag\\
&\hat{G}(\ell)=\frac{\Gamma[-\hat{\nu}]^2\,\Gamma[\sigma_++\hat{\phi}_+]\,\Gamma[\sigma_++\hat{\phi}_-]}{\Gamma[\hat{\nu}]^2\,\Gamma[\sigma_-+\hat{\phi}_+]\,\Gamma[\sigma_-+\hat{\phi}_-]},\notag\\
&\hat{\nu}=\sqrt{(1+2\,\ell)^2-4(1+\tilde{a}^2)(2+3\,\tilde{a}^2)\tilde{\mu}^2},\notag\\
&\sigma_\pm=\frac{1\pm\hat{\nu}}{2}+\hat{\delta},\quad\hat{\phi}_\pm=\pm\frac{i\,\tilde{a}\,\tilde{\mu}\,(m-\tilde{a}(1+\tilde{a}^2))}{1-\tilde{a}^2},\notag\\
&\hat{\delta}=\frac{i\,\xi}{1-\tilde{a}^2}\Big[1-\frac{\tilde{a}^4\,\tilde{\mu}^2\,(1-\tilde{a}^4)}{\xi^2}\Big]^{\frac{1}{2}},\notag\\
&\xi=\tilde{a}\,m-(1+\tilde{a}^2)\tilde{\mu},\label{eq:matching2}
\end{align}
For the imaginary part of $\tilde{\omega}$ we allow \eqref{eq:discreteCond} to be complex
\begin{align}
\frac{1}{2}+\nu-\kappa=-N+\epsilon,\label{eq:discreteCondPerturb}
\end{align}
with $\epsilon\ll1$, $\epsilon\in\mathbb{C}$. We can look at the $\epsilon\rightarrow0$ limit of \eqref{eq:discreteCondPerturb}, using \eqref{eq:realK}, to derive
\begin{align}
\operatorname{Im}\big(\sqrt{\tilde{k}}\big)=\epsilon\frac{2\,i\,e_1}{(1+2\,N+2\,\nu)^2},
\end{align}
which also enables us to find, in the limit $\ell\rightarrow\infty$,
\begin{align}
\tilde{\omega}_I=i\,\operatorname{Im}(\epsilon)\frac{(1+\tilde{a}^2)^2\,\tilde{\mu}^3}{4\,\ell^3}.\label{eq:imagFreqExpr}
\end{align}

Next, we look at \eqref{eq:matching2} for $\epsilon\rightarrow0$ and obtain
\begin{align}
(-1)^{N}\,N!\,\epsilon=\Big[\frac{\tilde{\mu}^2(1-\tilde{a}^4)}{N+\frac{1+\hat{\nu}}{2}}\Big]^{\hat{\nu}}\frac{\hat{G}(\ell)}{\Gamma[-N-\hat{\nu}]}.
\end{align}
Taking $\ell$ large, we can rearrange for $\epsilon$, which allows us to derive an expression for $\tilde{\omega}_I$ via \eqref{eq:imagFreqExpr}. It is valid for any spin parameter $\lvert a\rvert<M$ and scalar field mass $\mu$,
\begin{widetext}
	\begin{align}
	\tilde{\omega}_I&=\frac{2^{N-5-6\,\ell}(1+\tilde{a}^2)^{3+2\,\ell}\big((1-\tilde{a}^2)^2\ell^2+4\,\tilde{a}^2\,m^2\big)^{\frac{1}{2}+\ell}\tilde{\mu}^{5+4\,\ell}}{\sqrt{\pi}\,N!\,\ell^{-N+\frac{11}{2}+6\,\ell}}\,\sinh\bigg[\frac{2\,\pi\big(\tilde{a}\,m-(1+\tilde{a}^2)\tilde{\mu}\big)}{1-\tilde{a}^2}\bigg]\times\notag\\
	&\exp\bigg[2\,\ell-2(N+1)+\frac{2\,\tilde{\mu}\,(1+\tilde{a}^2)^2\arctan\big[\frac{2\,\tilde{a}\,m}{\ell(1-\tilde{a}^2)}\big]}{1-\tilde{a}^2}-\frac{4\,\tilde{a}\,m\,\arctan\big[\frac{2\,\tilde{a}\,m}{\ell(1-\tilde{a}^2)}\big]}{1-\tilde{a}^2}\bigg].\label{eq:omegaI}
	\end{align}
\end{widetext}
Setting $\ell=m$ ($n=0$) we get the growth rate of the dominant modes in the spectrum
\begin{widetext}
	\begin{align}
	\tilde{\omega}_{I,\ell=m}&=\frac{2^{N-5-6\,\ell}(1+\tilde{a}^2)^{4+4\,\ell}\tilde{\mu}^{5+4\,\ell}}{\sqrt{\pi}\,N!\,\ell^{-N+\frac{9}{2}+4\,\ell}}\sinh[\frac{2\,\pi(\tilde{a}\,\ell-\tilde{\mu}(1+\tilde{a}^2))}{1-\tilde{a}^2}]\times\notag\\
	&\exp\bigg[-2\bigg(N+1+\frac{\tilde{\mu}\,(1+\tilde{a}^2)^2\arctan\big[\frac{-2\,\tilde{a}}{1-\tilde{a}^2}\big]}{1-\tilde{a}^2}\bigg)+\ell\bigg(2+\frac{4\,\tilde{a}\,\arctan\big[\frac{-2\,\tilde{a}}{1-\tilde{a}^2}\big]}{1-\tilde{a}^2}\bigg)\bigg],
	\end{align}
\end{widetext}
which is rewritten in the main text for compactness. Moreover, for $\tilde{a}=0$ the $\sinh$ changes sign and we reproduce the correct behaviour for Schwarzschild \cite{futureWork}
\begin{align}
\tilde{\omega}_{I,\tilde{a}=0}=-\frac{2^{N-5-6\,\ell}e^{-2(N+1)+2\,\ell}\tilde{\mu}^{5+4\,\ell}}{\sqrt{\pi}\,N!\,\ell^{-N+\frac{9}{2}+4\,\ell}}\sinh[2\,\pi\,\tilde{\mu}].
\end{align}

{\bf~KG Numerically--} We apply spectral collocation methods on a discretised Chebyshev grid with
\begin{align}
z=\frac{1+\cos\theta}{2},\quad x=1-\frac{r_+}{r},\label{eq:numericalCoords}
\end{align}
where $x=0$ and $x=1$ correspond to the event horizon and spatial infinity, and $z=0$ and $z=1$ to the north and south poles, respectively. We use dimensionless variables \eqref{eq:dimlessVars}, factor out the singular behaviour at the boundaries
\begin{align}
&R_{\omega\ell m}(x)=(1-x)^\beta\,e^{\frac{\alpha}{1-x}}\,x^\gamma\,\mathscr{R}_{\omega\ell m}(x),\notag\\
&S_{\omega\ell m}(z)=z^{m/2}(1-z)^{m/2}\,\mathscr{S}_{\omega\ell m}(z),\notag\\
&\alpha=-\sqrt{\tilde{\mu}^2-\tilde{\omega}^2},\quad\beta=1+\frac{(1+\tilde{a}^2)(\tilde{\mu}^2-2\tilde{\omega}^2)}{2\sqrt{\tilde{\mu}^2-\tilde{\omega}^2}},\notag\\
&\gamma=-i\Big(\tilde{\omega}\frac{1+\tilde{a}^2}{1-\tilde{a}^2}-\frac{m\,\tilde{a}}{1-\tilde{a}^2}\Big),\label{eq:KerrQNMFactor}
\end{align}
and finally apply Newton's method. We then solve for $\mathscr{R}_{\omega\ell m}(x)$ and $\mathscr{S}_{\omega\ell m}(z)$.

No special treatment at the boundaries is needed as \eqref{eq:KerrQNMFactor} force the system to pick the right solution. An initial guess is \eqref{eq:omegaI}, $\Lambda=\ell(\ell+1)$, $\mathscr{R}_{\omega\ell m}(x)=1$, $\mathscr{S}_{\omega\ell m}(z)=1$.

\indent{\bf~Computation of $\psi_4$ --} We want to compute GWs emitted by a real scalar cloud around a Kerr BH. A complex scalar is not realistic and produces negligible gravitational radiation, since its stress-tensor is stationary - its use is a computational convenience for the determination of the field's QNMs, hence to find GWs we consider the real part of the superradiant modes - $\psi_R=\Re\,\psi$.

As explained in the main text, we treat the cloud as a perturbing source and use Teukolsky's equation \cite{Teukolsky:1973ha,Press:1973zz,Teukolsky:1974yv}
\begin{subequations}
	\begin{align}
	&\Big[\big(\Delta+3\gamma-\bar{\gamma}+4\mu+\bar{\mu}\big)\big(D+4\epsilon-\rho\big)-3\psi_2-\notag\\
	&\big(\bar{\delta}+3\alpha+\bar{\beta}+4\pi-\bar{\tau}\big)\big(\delta+4\beta-\tau\big)\Big]\psi_4=4\pi\,T_4,\label{eq:teukPsi4Supp}\\
	&T_4=\big(\Delta+3\gamma-\bar{\gamma}+4\mu+\bar{\mu}\big)\times\notag\\
	&\Big[\big(\bar{\delta}-2\bar{\tau}+2\alpha\big)T_{n\bar{m}}-\big(\Delta+2\gamma-2\bar{\gamma}+\bar{\mu}\big)T_{\bar{m}\bar{m}}\Big]\notag\\
	&+\big(\bar{\delta}-\bar{\tau}+\bar{\beta}+3\alpha+4\pi\big)\times\notag\\
	&\Big[\big(\Delta+2\gamma+2\bar{\mu}\big)T_{n\bar{m}}-\big(\bar{\delta}-\bar{\tau}+2\bar{\beta}+2\alpha\big)T_{nn}\Big].\label{eq:T4}
	\end{align}
\end{subequations}
to extract the NP scalar, $\psi_4$, encoding the information about the outgoing GWs. Our sign convention is $(-,+,+,+)$, opposite to Teukolsky's. Nevertheless, the equation in terms of NP variables is unchanged. For their definitions, in our convention, we use \cite{Frolov:1998wf}.

The tetrad projections we need in $T_4$ are $T_{nn}$, $T_{n\bar{m}}$ and $T_{\bar{m}\bar{m}}$. Moreover using a null tetrad implies we can ignore the $g_{\mu\nu}$ term. In a Kinnersley tetrad \cite{Kinnersley:1969zza} we have
\begin{align}
T_{nn}={}&\frac{e^{-2\,i\,\omega\,t}e^{2\,i\,m\,\phi}}{8\,\Sigma^2}\big[i\,K\,R_{\omega\ell m}+\Delta_k\,R_{\omega\ell m,r}\big]^2S_{\omega\ell m}^2,\notag\\
T_{n\bar{m}}={}&\frac{e^{-2\,i\,\omega\,t}e^{2\,i\,m\,\phi}}{4\sqrt{2}\,\Sigma\left(r-i\,a\,\cos\theta\right)}\big[i\,K\,R_{\omega\ell m}+\Delta_k\,R_{\omega\ell m,r}\big]\times\notag\\
&\bigg[S_{\omega\ell m,\theta}-\left(a\,\omega\,\sin\theta-\frac{m}{\sin\theta}\right)S_{\omega\ell m}\bigg]R_{\omega\ell m}\,S_{\omega\ell m},\notag\\
T_{\bar{m}\bar{m}}{}={}&\frac{e^{-2\,i\,\omega\,t}e^{2\,i\,m\,\phi}}{4\left(r-i\,a\,\cos\theta\right)^2}\,R_{\omega\ell m}^2\times\notag\\
&\bigg[S_{\omega\ell m,\theta}-\left(a\,\omega\,\sin\theta-\frac{m}{\sin\theta}\right)S_{\omega\ell m}\bigg]^2.
\end{align}

Furthermore, in the main text it is stated that the GW frequency and mode number are related to the scalar field ones by $\hat{\omega}=2\,\omega$, $\hat{m}=2\,m$. To see this we first remind ourselves of our ansatz for the scalar field
\begin{equation}
\psi(t,r,\theta,\phi)=e^{-i\,\omega\,t+i\,m\,\phi}R_{\omega\ell m}(r)S_{\omega\ell m}(\theta)\,.\label{eq:KGAnsatzSupp}
\end{equation}
Next, we define the stress tensor of a scalar as
\begin{equation}
T_{\mu\nu}\big(\Upsilon,\Psi\big)=2\nabla_\mu\Upsilon\nabla_\nu\Psi-g_{\mu\nu}\big(\nabla_\sigma\Upsilon\nabla^{\sigma}\Psi+\mu^2\Upsilon\Psi\big),\label{eq:TmunuDef}
\end{equation}
and plug in $\psi_R=1/2(\psi+\bar{\psi})$ for $\Upsilon$ and $\Psi$, leading us to
\begin{align}
T_{\mu\nu}\big(\psi_R,\psi_R\big)=\frac{1}{4}\Big[T_{\mu\nu}\big(\psi,\psi\big)+T_{\mu\nu}\big(\bar{\psi},\bar{\psi}\big)+2T_{\mu\nu}\big(\psi,\bar{\psi}\big)\Big],\notag
\end{align}
with bar indicating complex conjugation. The ansatz \eqref{eq:KGAnsatzSupp} implies that the terms above will contain exponential prefactors, which reveal that the first and second one source outgoing waves with $\hat{m}=2\,m$, $\hat{\omega}=2\,\omega$ and $\hat{m}=-2\,m$, $\hat{\omega}=-2\,\omega$, respectively, whereas the third one corresponds to the energy of the cloud, given by the gravitational mode with $\hat{m}=\hat{\omega}=0$.

The LHS of \eqref{eq:teukPsi4Supp} can be separated into angular and radial parts as in the vacuum case \cite{Teukolsky:1973ha}. Thus take
\begin{align}
\psi_4=e^{-i\,\hat{\omega}\,t}e^{i\,\hat{m}\,\phi}\,\rho^4\,\mathcal{R}(r)\,\mathcal{S}(\theta),\label{eq:psi4Sep}
\end{align}
with spin coefficient $\rho=-1/(r-i\,a\,\cos\theta)$, and multiply both sides by $2\,\Sigma(r,\theta)/\rho^4$. $\mathcal{S}(\theta)$ can be identified with a spin-weighted spheroidal harmonic, satisfying
\begin{align}
\frac{1}{\sin\theta}\Big[\sin\theta\,&\tensor[_s]{\mathcal{S}}{_{\hat{\ell}\hat{m},\theta}}\Big]_{,\theta}+\bigg[\left(c\,\cos\theta\right)^2-2\,c\,s\,\cos\theta+s\notag\\
&+\tensor[_s]{A}{_{\hat{\ell}\hat{m}\hat{\omega}}}-\frac{\left(\hat{m}+s\,\cos\theta\right)^2}{\sin\theta^2}\bigg]\tensor[_s]{\mathcal{S}}{_{\hat{\ell}\hat{m}}}=0,\label{eq:spheroidalODE}
\end{align}
with $c= a\,\hat{\omega}$, $\tensor[_s]{A}{_{\hat{\ell}\hat{m}\hat{\omega}}}$ the separation constant and $s=-2$. Multiplying by $\tensor[_s]{\bar{\mathcal{S}}}{_{\hat{\ell}\hat{m}}}$ and integrating over $\theta$ results in the following equation for the radial function
\begin{align}
\Delta_k^2\bigg[\Delta_k^{-1}\mathcal{R}_{\hat{\ell}\hat{m},r}\bigg]_{,r}+\bigg[\frac{K^2+4\,i\left(r-M\right)K}{\Delta_k}-8\,i\,\hat{\omega}\,r\notag\\
-a^2\,\hat{\omega}^2+2\,a\,\hat{m}\,\hat{\omega}-\tensor[_s]{A}{_{\hat{\ell}\hat{m}\hat{\omega}}}\bigg]\mathcal{R}_{\hat{\ell}\hat{m}}=\mathcal{T}_{\hat{\ell}\hat{m}\hat{\omega}},\label{eq:radialTeuk}
\end{align}
where $K=\left(r^2+a^2\right)\hat{\omega}-a\,\hat{m}$ and the source term is the integrated over angles stress-energy tensor
\begin{align}
\mathcal{T}_{\hat{\ell}\hat{m}\hat{\omega}}=\frac{4\pi}{\eta_{\hat{\ell}\hat{m}}}\int \frac{2\,\Sigma(r,\theta)}{\rho^4}\tensor[_s]{\bar{\mathcal{S}}}{_{\hat{\ell}\hat{m}}}\,T_4\,\sin\theta\,\mathrm{d}\theta,\label{eq:projectedSource}
\end{align}
with the normalisation condition
\begin{align}
\int_{0}^{\pi}\lvert\tensor[_s]{\mathcal{S}}{_{\hat{\ell}\hat{m}}}\left(\theta\right)\rvert^2\sin\theta\,\mathrm{d}\theta=\eta_{\hat{\ell}\hat{m}}.\label{eq:angularNorm}
\end{align}
Equation \eqref{eq:radialTeuk} should be evaluated for $\hat{\omega}=\pm2\,\omega$, $\hat{m}=\pm2\,m$, $s=-2$ on the LHS with $T_{\mu\nu}\big(\psi,\psi\big)$, for the $+$, and $T_{\mu\nu}\big(\bar{\psi},\bar{\psi}\big)$, for the $-$, inside $T_4$ on the RHS, and for $m=\hat{\omega}=0$ on the LHS with $T_{\mu\nu}\big(\psi,\bar{\psi}\big)$ in $T_4$ on the RHS.

However, the Teukolsky equation is invariant under complex conjugation followed by $m\rightarrow-m$ and $\omega\rightarrow-\omega$, hence for $\hat{m}\ne0\ne\hat{\omega}$ we only need to determine the contribution of $T_{\mu\nu}\big(\psi,\psi\big)$ and double the result.

Moreover, by looking at the asymptotic behaviour of the homogeneous radial equation (\eqref{eq:radialTeuk} with $\mathcal{T}_{\hat{\ell}\hat{m}\hat{\omega}}=0$) we can show that the $\hat{m}=\hat{\omega}=0$ mode is subleading to all the rest at spatial infinity. To this end, transform the homogeneous \eqref{eq:radialTeuk} with the help of
\begin{align}
\mathcal{R}_{\hat{\ell}\hat{m}}(r)=\frac{\Delta_K^{-\frac{s}{2}}}{\sqrt{r^2+a^2}}Y_{\hat{\ell}\hat{m}}(r),\quad\frac{dr_*}{dr}=\frac{r^2+a^2}{\Delta_K},
\end{align}
and then take the limit of $r\rightarrow\infty$, leaving us with
\begin{align}
Y_{\hat{\ell}\hat{m}}''(r_*)+\Big[\hat{\omega}^2+\frac{2\,i\,s\,\hat{\omega}}{r}\Big]Y_{\hat{\ell}\hat{m}}(r_*)=0,\label{eq:asympEq1}
\end{align}
for finite\footnote{Due to the aforementioned symmetry of the Teukolsky equation, we can work, without loss of generality, with the case of $\hat{\omega}=2\,\omega$, $\hat{m}=2\,m$ only.} frequency $\hat{\omega}$ and azimuthal number $\hat{m}$. The outgoing part of $Y_{\hat{\ell}\hat{m}}$ behaves as $e^{i\,r_*\,\hat{\omega}}r^{-s}$, leading to an $e^{i\,r_*\,\hat{\omega}}r^{-2s-1}$ asymptotic behaviour for $\mathcal{R}_{\hat{\ell}\hat{m}}$, which together with the definition \eqref{eq:psi4Sep} allows us to deduce that the outgoing contribution of $\psi_4$ near spatial infinity, for finite $\hat{\omega}$ and $\hat{m}$, behaves as
\begin{align}
\lim\limits_{r\rightarrow\infty}\psi_4\sim e^{i\,\hat{\omega}\,r_*}/r\quad({\rm outgoing\;mode}).
\end{align}
However, were we to set $\hat{\omega}=\hat{m}=0$ before we take the limit of $r\rightarrow\infty$, instead of \eqref{eq:asympEq1}, we are left with
\begin{align}
Y_{00}''(r_*)-\frac{\ell(\ell+1)}{r^2}Y_{00}(r_*)=0,
\end{align}
where we have used $\tensor[_s]{A}{_{\hat{\ell}00}}=\ell(\ell+1)-s(s+1)$. The solution with finite energy at infinity corresponds to asymptotic behaviour of the form $r^{-\ell-3}$ for $\psi_4$, which is clearly subleading to non-zero $\hat{\omega}$ and $\hat{m}$ modes.

Therefore, for the computation of the gravitational radiation from the scalar clouds, which is performed at spatial infinity, we only need to look at the $\hat{\omega}=2\,\omega$, $\hat{m}=2\,m$ case for \eqref{eq:radialTeuk}, with $T_{\mu\nu}\big(\psi,\psi\big)$ inside $T_4$ on the RHS.

In this way, $\psi_4$ has been projected onto a basis of spin-weighted spheroidal harmonics
\begin{align}
\psi_4=e^{-i\,\hat{\omega}\,t}e^{i\,\hat{m}\,\phi}\,\rho^4\,\sum_{\hat{\ell}}\mathcal{R}_{\hat{\ell}\hat{m}}(r)\,\tensor[_s]{\mathcal{S}}{_{\hat{\ell}\hat{m}}}(\theta),\label{eq:projectedPsi4}
\end{align}
where the definition of $\tensor[_s]{\mathcal{S}}{_{\hat{\ell}\hat{m}}}(\theta)$ requires $\hat{\ell}\geq\hat{m}$.

Thus, given a Kerr BH, for fixed $M\,\mu$ and $J/M^2$, and its $\ell=m$ scalar superradiant spectrum, we can use the latter as a source on the RHS of \eqref{eq:radialTeuk}, which sets the values of the GW frequency $\hat{\omega}$ and mode-number $\hat{m}$, letting us solve \eqref{eq:radialTeuk} for any allowed value of $\hat{\ell}$.

Knowing $\psi_4$, we can compute the gravitational radiation at infinity and the maximum of the curvature component represented by the NP scalar $\psi_4$ with the formulae given in the main text. The argument for ignoring the $\hat{m}=\hat{\omega}=0$ mode in the computation of the gravitational radiation at infinity does not hold for the maximum of $\psi_4$, as the latter is expected to be located at a finite distance from the black hole, near the peak of the scalar cloud, as we find in our numerics. Nevertheless, we have checked numerically that its contribution to the maximum is negligible compared to the rest.

\indent{\it~Numerical integration --} We use the same numerical method for \eqref{eq:spheroidalODE} and \eqref{eq:radialTeuk} as for \eqref{eq:KGrad} and \eqref{eq:KGang}.

\indent{\it~The angular equation --} \eqref{eq:spheroidalODE} is an eigenvalue problem which we integrate using Newton's method. We write
\begin{align}
\tensor[_s]{\mathcal{S}}{_{\ell m}}=z^{\iota_-}(1-z)^{\iota_+}\tensor[_s]{\mathbb{S}}{_{\ell m}},\label{eq:spinSpheroidalFactor}
\end{align}
where $\iota_{\pm}=\lvert\frac{m\pm s}{2}\rvert$ and solve for $\tensor[_s]{\mathbb{S}}{_{\ell m}}$. We drop the hats on $m$ and $\omega$, as we solve the equation generally. The eigenvalue $\tensor[_s]{A}{_{\ell m\omega}}$ is treated as an unknown and we normalise $\tensor[_s]{\mathbb{S}}{_{\ell m}}$ to $1$ using \eqref{eq:angularNorm}. Appropriate boundary conditions are imposed at the edges of the domain. These can be obtained by expanding the equation for $\tensor[_s]{\mathbb{S}}{_{\ell m}}$ in power series around the poles. At $z=0$ we require
\begin{align}
\big[-4+(m-s)^2&-8\,\iota_--(2\,\iota_-)^2\big]\left.\partial_z\tensor[_s]{\mathbb{S}}{_{\ell m}}(z)\right|_{z=0}=\notag\\
-\big[m^2-4\,s+2\,&m\,s-4\,\tensor[_s]{A}{_{\ell m}}-8\,\tilde{a}\,s\,\omega-4\,\tilde{a}^2\omega^2+\notag\\
(2\,\iota_-)^2+4\,\iota_+&+4\,\iota_-(1+2\,\iota_+)\big]\left.\tensor[_s]{\mathbb{S}}{_{\ell m}}(z)\right|_{z=1},
\end{align}
and similarly at $z=1$
\begin{align}
\big[-4+(m+s)^2&-8\,\iota_+-(2\,\iota_+)^2\big]\left.\partial_z\tensor[_s]{\mathbb{S}}{_{\ell m}}(z)\right|_{z=1}=\notag\\
\big[m^2-4\,s-2\,&m\,s-4\,\tensor[_s]{A}{_{\ell m}}+8\,\tilde{a}\,s\,\omega-4\,\tilde{a}^2\omega^2+\notag\\
(2\,\iota_+)^2+4\,\iota_+&+4\,\iota_-(1+2\,\iota_+)\big]\left.\tensor[_s]{\mathbb{S}}{_{\ell m}}(z)\right|_{z=1}.
\end{align}
A starting point to our iteration is the series solution, provided in \cite{Berti:2005gp}, up to twelfth order.

\indent{\it~The radial equation --} \eqref{eq:radialTeuk} is a sourced ODE, which we can invert once the RHS is known. As explained above, we need the superradiant modes of the scalar field on a fixed Kerr background, which we have numerically, so we can integrate \eqref{eq:radialTeuk} for any $\hat{\ell}$. We take out the prefactors in \eqref{eq:KerrQNMFactor} from the scalar field inside $T_4$, so that we can directly substitute our numerical solutions $\mathscr{R}_{\omega\ell m}(x)$ and $\mathscr{S}_{\omega\ell m}(z)$. We then factor out the singular behaviour of $\mathcal{R}_{\hat{\ell}\hat{m}}$ at infinity and the horizon, derived via Frobenius analysis, and transfer them to the RHS. Specifically
\begin{align}
\mathcal{R}_{\hat{\ell}\hat{m}}(x)=r_+^2(1-x)^{\hat{\beta}}\,e^{\frac{\hat{\alpha}}{1-x}}\,x^{2+2\,\hat{\gamma}}\,\mathbb{R}_{\hat{\ell}\hat{m}}(x),\label{eq:gravRadialPrefactors}
\end{align}
with $\mathbb{R}_{\hat{\ell}\hat{m}}(x)$ dimensionless and
\begin{align}
\hat{\alpha}=2\,i\,\tilde{\omega},\;\hat{\beta}=-3-2\,i(1+\tilde{a}^2)\,\tilde{\omega},\notag\\
\hat{\gamma}=-i\left(\tilde{\omega}\frac{1+\tilde{a}^2}{1-\tilde{a}^2}-\frac{\hat{m}\,\tilde{a}}{1-\tilde{a}^2}\right).
\end{align}
Moreover, $x^{2\,\hat{\gamma}}$ gets cancelled by the $x^\gamma$ from the radial scalar functions \eqref{eq:KerrQNMFactor} inside $T_4$. We then solve for $\mathbb{R}_{\hat{\ell}\hat{m}}(x)$.
We need to impose boundary conditions at spatial infinity, since \cite{Teukolsky:1973ha} reveals that the ingoing behaviour of the radial function (the one we do not want) is subleading
\begin{align}
\mathcal{R}_{\hat{\ell}\hat{m}}(x)\sim Z^{(in)}_{\hat{\ell}\hat{m}}\,e^{-\frac{2\,i\,\tilde{\omega}}{1-x}}(1-x)+Z^{(out)}_{\hat{\ell}\hat{m}}\frac{e^{\frac{2\,i\,\tilde{\omega}}{1-x}}}{(1-x)^3},
\end{align}
The boundary condition that selects the outgoing waves only can be deduced by expanding the homogeneous Teukolsky equation for large radial variable , which shows that setting $Z^{(in)}_{\hat{\ell}\hat{m}}=0$ is equivalent to demanding
\begin{align}
&\partial_{xx}\left.\mathbb{R}_{\hat{\ell}\hat{m}}(x)\right|_{x=1}+\frac{1}{4(1-\tilde{a}^2)^2\hat{\varpi}^2}\Big[(1-\tilde{a}^2)^2\tensor[_s]{A}{_{\hat{\ell}\hat{m}\hat{\varpi}}}^2\notag\\
&+2(1-\tilde{a}^2)\tensor[_s]{A}{_{\hat{\ell}\hat{m}\hat{\varpi}}}\big[]2\,\tilde{a}\,\hat{m}\,\hat{\varpi}+\tilde{a}^4\hat{\varpi}^2+2\,\tilde{a}^6\hat{\varpi}^2\notag\\
&-(i+\hat{\varpi})^2+\tilde{a}^2(-1+4\,i\,\hat{\varpi}-3\,\hat{\varpi}^2)\big)\notag\\
&+2\,\hat{\varpi}\big(-3\,i-12\,\hat{\varpi}+20\,i\,\hat{\varpi}^2+\tilde{a}^8(8\,i-\frac{11}{2}\hat{\varpi})\hat{\varpi}^2\notag\\
&+8\,\hat{\varpi}^3+2\,\tilde{a}^{12}\,\hat{\varpi}^3+2\,\tilde{a}^{10}\hat{\varpi}^2(i+\hat{\varpi}-2\,\tilde{a}^7\,\hat{m}\,\hat{\varpi}(i-2\,\hat{\varpi})\notag\\
&+4\,\tilde{a}^5\,\hat{m}(1+\frac{i}{2}\hat{\varpi}+\frac{1}{2}\hat{\varpi}^2)-8\,\tilde{a}^3\,\hat{m}(1-\frac{3\,i}{2}\hat{\varpi}+\frac{3}{4}\hat{\varpi}^2)\notag\\
&+4\,\tilde{a}\,\hat{m}(1-3\,i\,\hat{\varpi}-2\,\hat{\varpi}^2)+\tilde{a}^4(3\,i-2\,\hat{\varpi}-20\,i\,\hat{\varpi}^2+\frac{1}{2}\hat{\varpi}^3)\notag\\
&+\tilde{a}^2(3\,i+19\,\hat{\varpi}+2\,\hat{m}^2\,\hat{\varpi}-4\,i\,\hat{\varpi}^2+12\,\hat{\varpi}^3)\notag\\
&-\tilde{a}^6(3\,i+5\,\hat{\varpi}+6\,i\,\hat{\varpi}^2+11\,\hat{\varpi}^3)\big]\Big]\mathbb{R}_{\hat{\ell}\hat{m}}(x=1)=0,
\end{align}
where $\hat{\varpi}=2\,\tilde{\omega}$.

In terms of the functions for which we solve numerically, the formula for the gravitational radiation given in the main text takes the following form
\begin{align}
\frac{dE_{s}}{dt}=\frac{1}{(2\,\tilde{\omega})^2}\sum_{\hat{\ell}}\eta_{\hat{\ell}\hat{m}}\left\lvert \hat{\mathbb{R}}_{\hat{\ell}\hat{m}}(x)\right\rvert^2_{x\rightarrow1},\label{eq:energyRateNumSupp}
\end{align}
where we have used \eqref{eq:angularNorm}, the orthogonality of the spin-weighted spheroidal harmonics, and have multiplied by 2 to accound for the GWs with $\hat{m}=-2\,m$, $\hat{\omega}=-2\,\omega$.

\begin{figure}
	\centering
	\includegraphics[width=.47\textwidth]{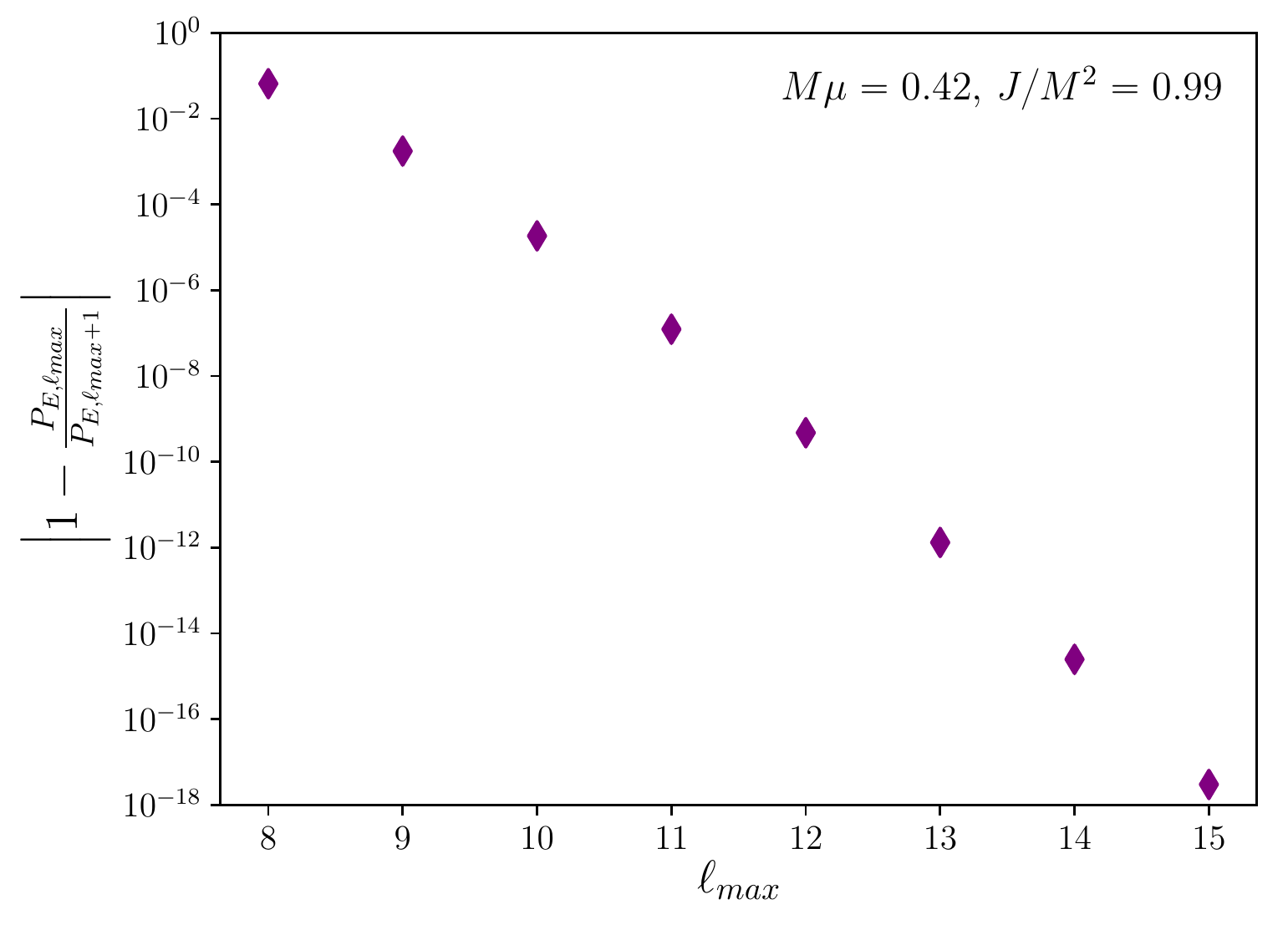}
	\caption{Convergence of the radiated GW energy at infinity $P_E$ for the $m=4$ scalar cloud as a function of the number of included $\ell$ modes in the projection of $\psi_4$ onto the basis of spin-weighted spheroidal harmonics at the highest grid resolution}
	\label{fig:GWemissionlModesConvergence}
\end{figure}

\begin{figure}
	\centering
	\includegraphics[width=.47\textwidth]{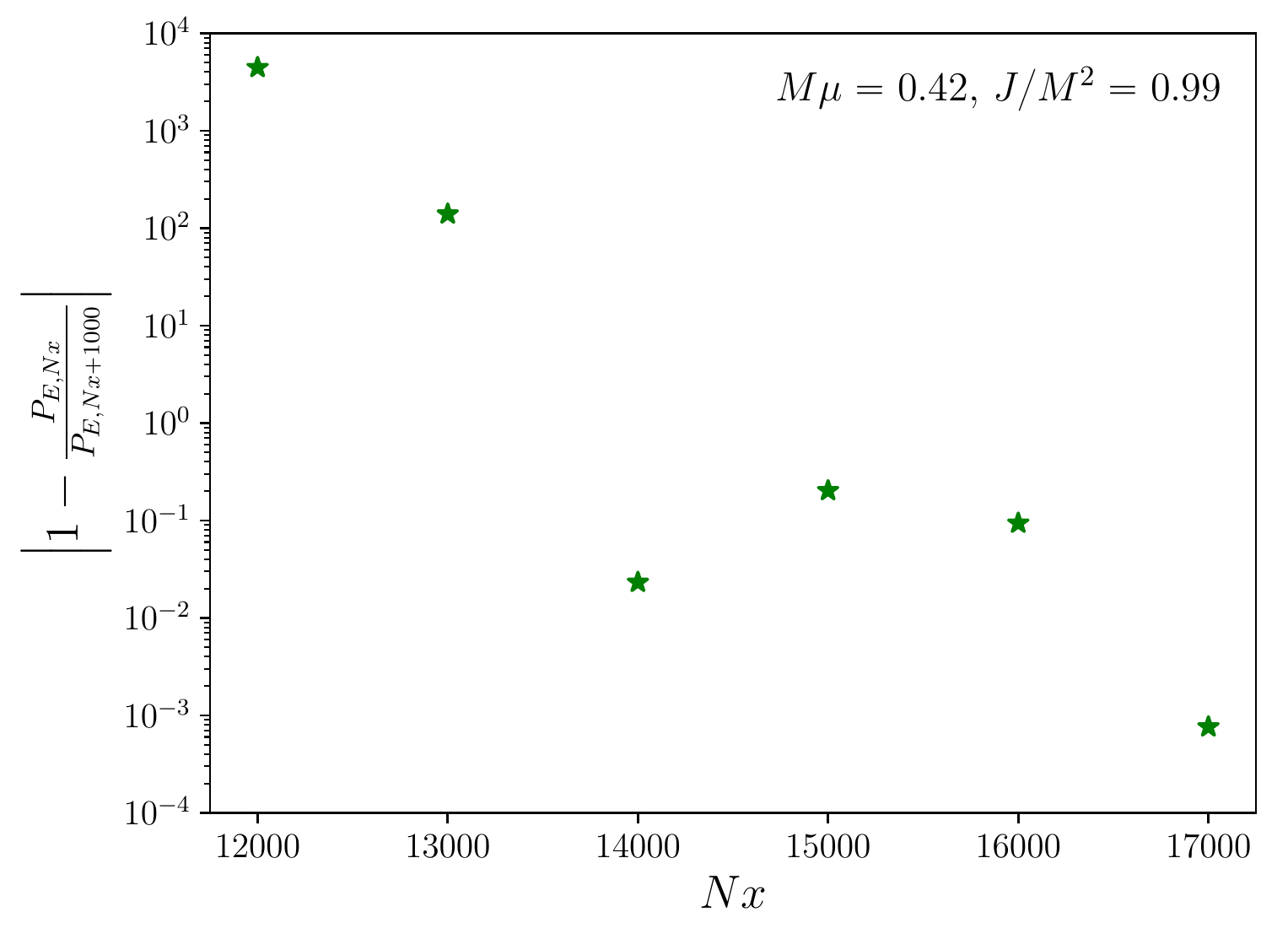}
	\caption{Convergence of the radiated GW energy at infinity $P_E$ for the $m=4$ scalar cloud as a function of the number of points in the radial direction $Nx$}
	\label{fig:GWemissionConvergence}
\end{figure}

\indent{\bf~Numerical convergence --} As we cannot integrate \eqref{eq:radialTeuk} for infinitely many $\hat{\ell}$, we truncuate the sum until \eqref{eq:energyRateNumSupp} converges. We will look at the $m=4$ case here ($\hat{m}=8$ for the GWs), as this was the hardest one to tackle numerically. We include 9 GW modes, $\hat{\ell}=8$ to $\hat{\ell}=16$, to get good convergence for the radiated energy. Moreover, very high grid resolution was needed in the radial direction, in order to resolve the oscillating behaviour of the solution far away from the BH (exactly at spatial infinity the oscillating part is discarded by the boundary conditions). This is summarised in Figs.~\eqref{fig:GWemissionlModesConvergence} and \eqref{fig:GWemissionConvergence}

\indent{\bf~Analytic approximation to $\chi$ --} In this section we outline the steps for the WKB approximation of $\chi \equiv \max_{r,\theta}\;\left(|\psi_4|^2/M_{s}^2\right)$. We work with $\ell=m$ and $\ell\gg1$. 

First, we need a WKB expansion of the scalar field near its maximum - i.e. $R_{\omega\ell m}(r)$, $S_{\omega\ell m}(\theta)$, $\Lambda$ and $\tilde{\omega}$, as well as the location of the minimum of the potential of the radial KG equation expanded for large $\ell=m$.

We start with the angular equation \eqref{eq:KGang}, which rewriten in our numerical coordinates \eqref{eq:numericalCoords}, together with the redefinition \eqref{eq:KerrQNMFactor}, looks like
\begin{align}
&(1-z)\,z\,\mathscr{S}_{\omega\ell m}''(z)+(m+1)(2z-1)\mathscr{S}_{\omega\ell m}'(z)\notag\\
&-\big[\Lambda-m(m+1)-\tilde{a}^2\,\tilde{k}\,(2z-1)^2\big]\mathscr{S}_{\omega\ell m}(z)=0.
\end{align}
Expand the variables in series in $m$
\begin{align}
\mathscr{S}_{\omega\ell m}(z)&=1+\frac{S_1(z)}{m}+\frac{S_2(z)}{m^2}+\mathcal{O}(m^{-3}),\notag\\
\Lambda&=\lambda_1\,m^2+\lambda_2\,m+\lambda_3+\mathcal{O}(m^{-1}),\notag\\
\tilde{\omega}&=\tilde{\mu}-\frac{(1+\tilde{a}^2)^2\,\tilde{\mu}^3}{8\,m^2}+\frac{\omega_3}{m^3}+\mathcal{O}(m^{-4}),
\end{align}
plug them in and solve order by order, requiring regularity of the solution. The known behaviour of $\tilde{\omega}$ from before is utilised. In this way one can obtain $S_i(z)$ and $\lambda_i$ to any order as a function of the $\omega_i$. 

Afterwards we transform the radial part \eqref{eq:KGrad} using 
\begin{align}
r=\frac{2\,r_+\,m(m+1)}{(1+\tilde{a}^2)\tilde{\mu}^2}u.\label{eq:coordNearMax}
\end{align}
The equation will not be given here - the above transformation correponds to the first two terms in the expansion for large $\ell=m$ of the position of the minimum of the radial potential. This is straightforward to get after deriving the leading order behaviour of the angular eigenvalue $\Lambda=m(m+1)+\mathcal{O}(1)$, for which no knowledge of $\omega_3$ or higher is required. Afterwards, we take
\begin{align}
R_{\omega\ell m}(u)=e^{m\,f(u)}\Big[1+\frac{R_1(u)}{m}+\mathcal{O}(m^{-2})\Big],
\end{align}
and solve order by order for $f(u)$, $R_i(u)$ and $\omega_i$, again demanding regularity of the solution. This can be done to any order, but we only give the first few here
\begin{align}
S_1(z)&=c_2,\quad S_2(z)=c_3,\notag\\
f(u)&=c_0-u+\log\,u,\notag\\
R_1(u)&=c_1-\frac{\tilde{\mu}^2}{4}(1+\tilde{a}^2)^2\big[u+\frac{1}{u}+3 \log\,u\big],\notag\\
\lambda_1&=\lambda_2=1,\quad\lambda_3=0,\quad\omega_3=\frac{(1+\tilde{a}^2)^2\,\tilde{\mu}^3}{4},
\end{align}
whereby $c_i$ are constants that can for example be chosen so that the scalar field is unity at the maximum ($u=1$), which is what we do to compare with the numerics (where we can just divide by the numerical maximum).

Next, we move onto the Teukolsky equation \eqref{eq:radialTeuk}. The expressions are very lengthy and not illuminating at all, so we will give only the most important ones. Treating the angular equation \eqref{eq:spheroidalODE} as the scalar one produces
\begin{align}
\tensor[_s]{\mathcal{S}}{_{\hat{\ell}\hat{m}}}&=z^{\lvert\frac{2m-s}{2}\rvert}(1-z)^{\lvert\frac{2m+s}{2}\rvert}\Big[1-\notag\\
&-\frac{2\,\tilde{a}\,\tilde{\mu}\,z(s+\tilde{a}\,\tilde{\mu}\,(1-z))+d_1}{m}+\mathcal{O}(m^{-2})\Big],\notag\\
\tensor[_s]{A}{_{\hat{\ell}\hat{m}\hat{\omega}}}&=2\,m\,(2\,m+1)-s(s+1)+\mathcal{O}(m^{-1}),\label{eq:spheroidalWKB}
\end{align}
whereby we have accounted for $\hat{m}=2\,m$ and $d_1$ is a constant that can be chosen to make the normalisation of the function easier\footnote{Depending on how one does the numerics, $\tensor[_s]{\bar{\mathcal{S}}}{_{\hat{\ell}\hat{m}}}$ inside $T_4$ might also need to be normalised.}. Afterwards we need to obtain a series expansion in $\ell=m$ for $\mathcal{T}_{\hat{\ell}\hat{m}\hat{\omega}}$ \eqref{eq:projectedSource}. The normalisation $\eta_{\hat{\ell}\hat{m}}$ has to be obtained by plugging \eqref{eq:spheroidalWKB} in \eqref{eq:angularNorm} and integrating. The final result is
\begin{align}
&\mathcal{T}_{\hat{\ell}\hat{m}\hat{\omega}}=e^{2\,m(1-u+\log\,u)}\frac{16\,\sqrt{2}\,\pi\,r_+^2\,u^4}{(1+\tilde{a}^2)^4\,\tilde{\mu}^6}m^9\Bigg[1-\frac{4}{m}\notag\\
&-\frac{\tilde{\mu}^2 \big[3\,(1+\tilde{a}^4)(u-1)^2+\tilde{a}^2(u\,(6\,u-13)+6)\big]}{6\,u\,m}\notag\\
-&\frac{2\,i\,\tilde{\mu}(1+\tilde{a}^2)+3\,\tilde{\mu}^2(1+\tilde{a}^2)^2\log (u)}{2\,m}+\mathcal{O}\Big[\frac{1}{m^2}\Big]\Bigg].
\end{align}
With all ingredients present, we change variables in \eqref{eq:radialTeuk} with \eqref{eq:coordNearMax} and WKB expand $\mathcal{R}_{\hat{\ell}\hat{m}}$ for large $\ell=m$. Matching order by order resutls in
\begin{align}
&\mathcal{R}_{\hat{\ell}\hat{m}}=-e^{2\,m(1-u+\log\,u)}\frac{\sqrt{2}\,\pi\,r_+^2\,u^2}{(1+\tilde{a}^2)^2\,\tilde{\mu}^4}m^5\Bigg[1-\frac{2}{m}\notag\\
&-\frac{\tilde{\mu}^2 \big[3\,(1+\tilde{a}^4)(u-1)^2+\tilde{a}^2(u\,(6\,u-13)+6)\big]}{6\,u\,m}\notag\\
-&\frac{2\,i\,\tilde{\mu}(1+\tilde{a}^2)+3\,\tilde{\mu}^2(1+\tilde{a}^2)^2\log (u)}{2\,m}+\mathcal{O}\Big[\frac{1}{m^2}\Big]\Bigg],
\end{align}
allowing us to to rebuild $\psi_4$ for $\ell=m$ via \eqref{eq:projectedPsi4}.

We also need an expression for the scalar cloud energy
\begin{equation}
\mathcal{M}_s=\int_{r_+}^{+\infty}\int_{0}^{\pi}\int_{0}^{2\pi}\sqrt{-g}\,\tensor{T}{^t_t}\,\mathrm{d}\phi\,\mathrm{d}\theta\,\mathrm{d}r.
\end{equation}
Working with coordinates $z$ and $u$, we substitute our expansions for the scalar field in the definition of $\mathcal{M}_s$ and after simplifications the angular integrals can be done analytically - with \textit{Mathematica} or with a book on integrals. The resulting expression is expanded in $\ell=m$ and integrated over $u$ from $\frac{\tilde{\mu}^2(1+\tilde{a}^2)}{2\,m\,(m+1)}$ to infinity, producing
\begin{align}
\mathcal{M}_s=\frac{e^{2\,m}\pi^{\frac{3}{2}}\,\Gamma\Big[3+2m,\frac{(1+\tilde{a}^2)\tilde{\mu}^2}{m+1}\Big]}{2^{4\,m-1}m^{2\,m-\frac{5}{2}}(1+\tilde{a}^2)^3\,\tilde{\mu}^4}\Big[1+\mathcal{O}(m^{-1})\Big],
\end{align}
where we also have higher order terms, but the expressiosn are too cumbersome.

Having $\psi_4$ and $\mathcal{M}_s$ in the large $\ell=m$ limit allows one to derive the expression for $\chi_{WKB}$ in the main text by dividing them and expanding in series in $\ell=m$.
\bibliography{Paperbib}
\bibliographystyle{unsrt}
\end{document}